\documentclass[twocolumn]{aastex701}

\usepackage[super]{nth}
\usepackage{hyperref}
\usepackage{booktabs}
\usepackage{comment}
\usepackage{listings}
\usepackage{color}
\usepackage{threeparttable}
\usepackage{url}
\usepackage{adjustbox}
\usepackage{algorithm}
\usepackage{algorithmic}
\usepackage{amsmath}

\newcommand{\ticstar}{TIC~295741342}
\newcommand{\kms}{\ifmmode{\rm km\thinspace s^{-1}}\else km\thinspace s$^{-1}$\fi}

\usepackage{xcolor}

\newcommand{\rgbmaa}{1.08~}
\newcommand{\rgbmab}{1.04~}
\newcommand{\rgbmb}{1.70~}
\newcommand{\rgbraa}{1.06~}
\newcommand{\rgbrab}{1.01~}
\newcommand{\rgbrb}{10.61~}
\newcommand{\rgbpa}{4.75}
\newcommand{\rgbpab}{412.8}

\newcommand{\rgbeab}{0.363~}

\newcommand{\rgbmutinc}{0.33~}
\newcommand{\rgbmutinctab}{0.3~}
\newcommand{\rgbteffb}{4839~}
\newcommand{\rgbloggb}{2.62~}
\newcommand{\rgbageb}{1.46~}
\newcommand{\rgbrlof}{54~}
\newcommand{\rgbrlofrad}{83~}

\newcommand{\hbmaa}{1.11~}
\newcommand{\hbmab}{1.07~}
\newcommand{\hbmb}{1.81~}
\newcommand{\hbraa}{1.09~}
\newcommand{\hbrab}{1.05~}
\newcommand{\hbrb}{10.64~}
\newcommand{\hbpa}{4.75}
\newcommand{\hbpab}{412.8}

\newcommand{\hbeab}{0.362~}

\newcommand{\hbmutinc}{0.25~}
\newcommand{\hbmutinctab}{0.2~}
\newcommand{\hbteffb}{4991~}
\newcommand{\hbloggb}{2.64~}
\newcommand{\hbageb}{1.24~}
\newcommand{\hbrlof}{129~}
\newcommand{\hbrlofrad}{85~}

\begin{document}

\title{TIC 295741342: A Triply-Eclipsing Triple Star System with a Giant Tertiary}

\correspondingauthor{Brian P. Powell}
\email{brian.p.powell@nasa.gov}

\author[0000-0003-0501-2636]{Brian P. Powell}
\affiliation{NASA Goddard Space Flight Center, 8800 Greenbelt Road, Greenbelt, MD 20771, USA}
\email{brian.p.powell@nasa.gov}
\author[0000-0002-5286-0251]{Guillermo Torres}
\affiliation{Center for Astrophysics $\vert$ Harvard \& Smithsonian, 60 Garden Street, Cambridge, MA 02138, USA}
\email{gtorres@cfa.harvard.edu}
\author[0000-0001-9786-1031]{Veselin~B.~Kostov}
\affiliation{NASA Goddard Space Flight Center, 8800 Greenbelt Road, Greenbelt, MD 20771, USA}
\affiliation{SETI Institute, 189 Bernardo Ave, Suite 200, Mountain View, CA 94043, USA}
\email{veselin.b.kostov@nasa.gov}

\author[0000-0003-3182-5569]{Saul A. Rappaport}
\affiliation{Department of Physics, Kavli Institute for Astrophysics and Space Research, M.I.T., Cambridge, MA 02139, USA}
\email{sar@mit.edu}

\author[0000-0002-8806-496X]{Tam\'as Borkovits}
\affiliation{Baja Astronomical Observatory of University of Szeged, H-6500 Baja, Szegedi út, Kt. 766, Hungary}
\affiliation{HUN-REN -- SZTE Stellar Astrophysics Research Group,  H-6500 Baja, Szegedi út, Kt. 766, Hungary}
\affiliation{Konkoly Observatory, Research Centre for Astronomy and Earth Sciences, H-1121 Budapest, Konkoly Thege Miklós út 15-17, Hungary}
\email{borko@bajaobs.hu}

\author[0000-0002-5665-1879]{Robert Gagliano}
\affiliation{Amateur Astronomer, Glendale, AZ 85308}
\email{astrowebdoc@gmail.com}

\author[0000-0001-9911-7388]{David W. Latham}
\affiliation{Center for Astrophysics $\vert$ Harvard \& Smithsonian, 60 Garden Street, Cambridge, MA 02138, USA}
\email{dlatham@cfa.harvard.edu}

\begin{abstract}
We present the discovery and characterization of TIC 295741342, a triply-eclipsing triple star system with a giant tertiary. The eclipsing binary consists of two similar main-sequence stars in a \hbpa-day orbit.  The binary is in a \hbpab-day orbit with the giant tertiary.  We found two degenerate solutions for the system: one where the tertiary is ascending the Red Giant Branch (RGB), and the other where the tertiary is on the Horizontal Branch (HB) and will eventually ascend the Asymptotic Giant Branch (AGB).  In both solutions, the system is near-perfectly coplanar. In {\em TESS} Sector 33, the binary passes behind the giant tertiary, producing a distinctive ``head-and-shoulders'' eclipse that directly constrains the relative flux contributions and radii of all three stars. We modeled the system using a comprehensive spectro-photodynamical model that simultaneously fits the {\em TESS} lightcurve, eclipse times, spectral energy distribution, and radial velocities from 48 TRES spectra obtained over four years of observation resolving all three components. Evolutionary analysis using MIST tracks indicates that, in both solutions, the tertiary will overflow its Roche lobe, one in the RGB and the other in the AGB.  The Roche lobe overflow will initiate either a stable mass transfer to the binary or a common envelope evolution that will likely result in ejections and/or mergers. Our models predict the midpoint of the next outer eclipse will occur on September 1, 2026 and we encourage follow-up observations with a $\pm$3 day window to observe the full event and further constrain the system parameters.
\end{abstract}

\keywords{stars: binaries (including multiple): close - stars: binaries: eclipsing}


\section{Introduction}
\label{sec::intro}

Triple star systems, particularly those that are triply eclipsing, have seen rapid growth in discovery and study in the eras of the {\em Kepler} \citep{2010Sci...327..977B} and Transiting Exoplanet Survey Satellite \citep[{\em TESS};][]{Ricker14} missions.  In the {\em Kepler} era, the discoveries of KOI-126 by \citet{2011Sci...331..562C} and HD 181068 by \citet{2011Sci...332..216D} were followed by additional triply-eclipsing triples discovered by \citet{2013ApJ...768...33R}, while eclipse timing variations (ETVs) and eclipse depth variations proved to be powerful tools for characterizing triples, as demonstrated by \citet{2016MNRAS.455.4136B} and \citet{2022MNRAS.515.3773B}. A comprehensive study of the dynamical interactions within compact hierarchical triples was provided by \citet{2022MNRAS.510.1352B}. 

The breadth of the {\em TESS} field, now covering nearly the entirety of the sky, has enabled the discovery of many more triply-eclipsing triples. Early {\em TESS} discoveries were reported by \citet{Borkovits2020}, \citet{2020MNRAS.496.4624B}, and \citet{2020MNRAS.498.6034M}. Many more have followed in systematic studies \citep{2022MNRAS.513.4341R,2023MNRAS.521..558R,2024A&A...686A..27R}. \citet{2024A&A...685A..43M} identified 125 new triples through long-term ETVs in the Northern Continuous Viewing Zone, and \citet{2025A&A...695A.209B} revisited the {\em Kepler} triples of \citet{2016MNRAS.455.4136B} with new {\em TESS} observations.

Complementary spectroscopic and astrometric approaches have further expanded the census of known triples. Systematic radial velocity surveys by \citet{2020AJ....160..251T,2022ApJ...926....1T,2023AJ....165..165T,2025AJ....169..124T} have confirmed and resolved the orbits of numerous triples, while {\em Gaia} DR3 data \citep{DR3} has enabled the identification of hundreds of triple candidates through orbital solutions \citep{2023MNRAS.526.2830C,Czavalinga2023}, combined analyses with known {\em TESS} eclipsing binaries \citep{2023MNRAS.524.4296P}, and radial velocity--astrometric cross-matching \citep{2024A&A...692A.247B}.

Perhaps the most stunning recent development in the study of triple star systems was the discovery of TIC 290061484 by \citet{2024ApJ...974...25K}. This triple star system has an outer period of just 24.5 days and, as the authors appropriately stated, `eclipses' the previous record of 30.5 days held by Lambda Tauri \citep{1956ApJ...124..507E} since 1956.\footnote{Recently, \citet{2026arXiv260420314K} suggested that the eclipsing binary V0885 Per is part of a 21.8 day triple star system.  No third body eclipses are seen, and the conclusion is based on radial velocity measurements. If confirmed, this system would become the new shortest period triple known.} As the field continues to produce such unexpected systems, each new discovery offers fresh constraints on the formation and evolution of stellar multiples.

In this paper, we add to this growing body of work by presenting the discovery of TIC 295741342, a near-perfectly coplanar triply-eclipsing triple star system with a giant tertiary star. This paper is organized as follows.  In Section \ref{sec::tess}, we will discuss the {\em TESS} observations of TIC 295741342, including the outer eclipse that first alerted us to this system.  In Section \ref{sec::spectroscopy}, we will detail the spectroscopic observations and RVs.  In Section \ref{sec::model}, we will discuss the various aspects of our system model, followed by the details of the solution in Section \ref{sec::results}. In Section \ref{sec::discussion}, we will further analyze the system solution, to include the prediction of future eclipses and evolutionary changes.  Finally, in Section \ref{sec::summary}, we will summarize our study of this system.

\section{{\em TESS} Observations}
\label{sec::tess}

\begin{figure}
   \centering
    \includegraphics[width=1.\columnwidth]{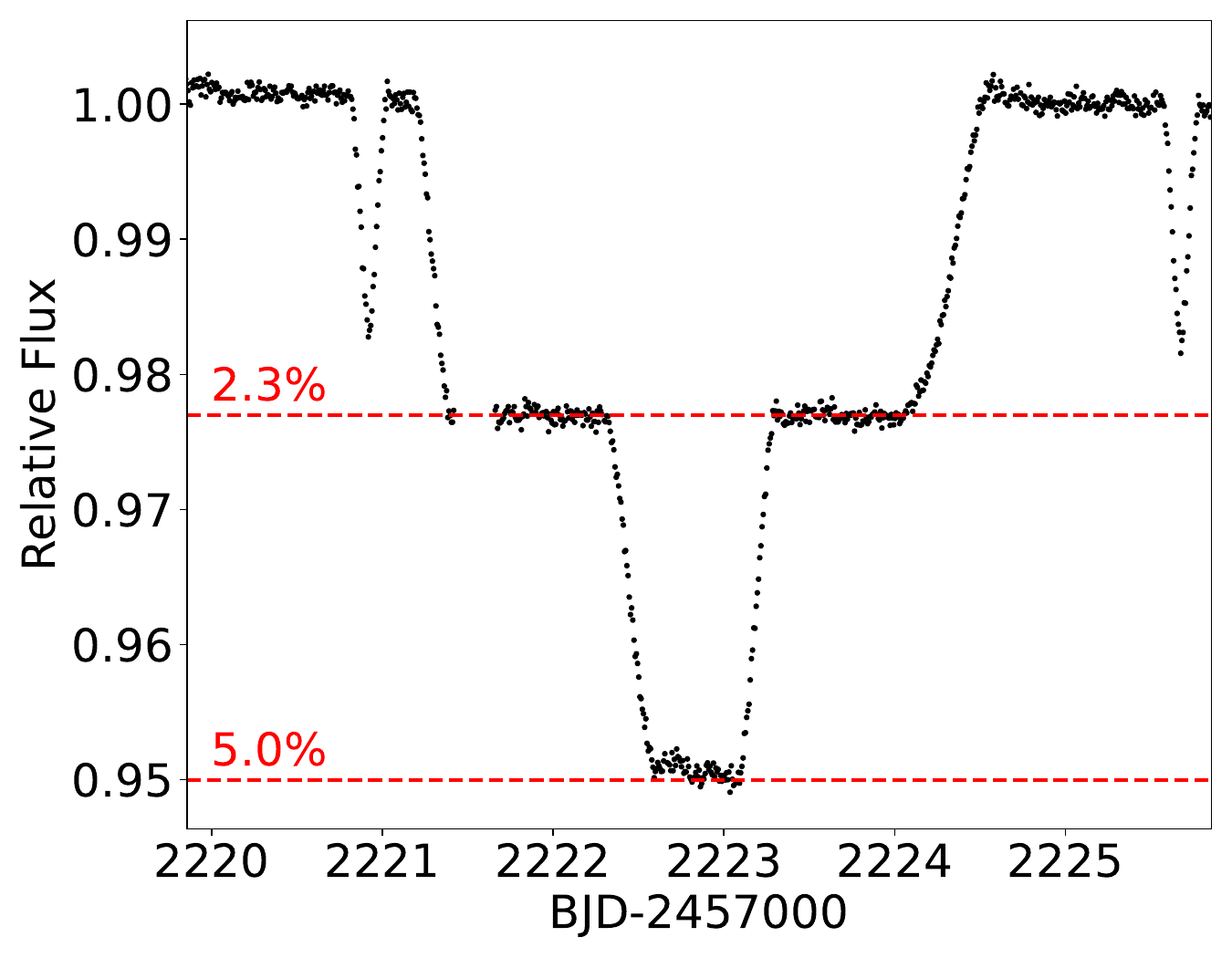}
    \caption{TIC 295741342 outer-body eclipse.  The {\em TESS} flux is shown in black points, with horizontal dashed red lines indicating the depths of the `shoulders' and `head' of the outer body eclipse, which substantially constrains the relative fluxes of the stars in the system in the {\em TESS} band.}
    \label{fig::zoom}
\end{figure}

{\em TESS} observed TIC 295741342 (stellar parameters shown in Table \ref{tab:stellar_params}) in Sectors 7, 33, 34, and 87.  For this study, we used lightcurves from the MIT Quick Look Pipeline \citep[QLP;][]{2020RNAAS...4..204H,2020RNAAS...4..206H,2021RNAAS...5..234K,2022RNAAS...6..236K}.  We provide a zoomed-in view of the Sector 33 outer-body eclipse in Figure \ref{fig::zoom} with a horizontal red line at 2.3\% corresponding to the flat ``shoulders'' of the eclipse, and another horizontal red line at 5\% corresponding to the depth of the ``head.''  The shape of the eclipse demonstrates the secondary of the EB passing fully behind a larger star (first shoulder), followed by the primary together with the secondary (the head), followed by the primary emerging from behind the tertiary (second shoulder).  From a simple visual analysis of Figure \ref{fig::zoom}, the shape of the eclipse constrains relative fluxes in the {\em TESS} band to the secondary to contributing 2.3\% of the system light and the primary to 2.7\% of the system light, with the larger outer body providing 95\% of the system light.  We will provide details of the mechanics of this eclipse in Section \ref{sec::results}.  

\begin{deluxetable*}{l r r r}[!ht]
\tabletypesize{\scriptsize}
\tablecaption{Basic parameters for TIC 295741342\label{tab:stellar_params}}
\tablewidth{0pt}
\tablehead{
\colhead{Parameter} & \colhead{Value} & \colhead{$\sigma$} & \colhead{Source}
}
\startdata
\multicolumn{4}{l}{\bf Identifying Information} \\
\hline
TIC ID & 295741342 &  & TIC\\
Gaia DR3 ID & 3032920469230415360 &  & Gaia DR3\\
Tycho-2 & 5406-00497-1 &  & TIC\\
2MASS & 07154586-1329070 &  & TIC\\
AllWISE & J071545.86-132907.0 &  & TIC\\
$\alpha$ (J2000, degrees) & 108.9411 &  & TIC\\
$\delta$ (J2000, degrees) & -13.4853 &  & TIC\\
$l$ (degrees) & 227.6786 &  & TIC\\
$b$ (degrees) & -0.8390 &  & TIC\\
\\
\multicolumn{4}{l}{\bf Gaia Measurements} \\
\hline
$\mu_{\alpha}$ (mas~yr$^{-1}$) & -5.9103 & 0.0410 & Gaia DR3\\
$\mu_{\delta}$ (mas~yr$^{-1}$) & 7.7306 & 0.0484 & Gaia DR3\\
$\varpi$ (mas) & 1.2051 & 0.0399 & Gaia DR3\\
RUWE & 2.81 &  & Gaia DR3\\
\tt{astrometric\_excess\_noise} & 0.33 &  & Gaia DR3\\
\tt{astrometric\_excess\_noise\_sig} & 175.08 &  & Gaia DR3\\
\\
\multicolumn{4}{l}{\bf Derived Stellar Properties} \\
\hline
$T_{\rm eff}$ (K) & 5085 & 122 & TIC\\
$R$ ($R_\odot$) & 11.1774 & -- & TIC\\
Luminosity class & GIANT &  & TIC\\
$d$ (pc) & 945.2 & $^{+59.8}_{-53.2}$ & TIC\\
$E(B-V)$ (mag) & 0.1090 & $^{+0.0247}_{-0.0174}$ & TIC\\
\\
\multicolumn{4}{l}{\bf Photometric Properties} \\
\hline
$T$ (mag) & 9.6280 & 0.0141 & TIC\\
$G$ (mag) & 10.2549 & 0.0002 & Gaia DR3\\
$G_{\rm BP}$ (mag) & 10.7773 & 0.0003 & Gaia DR3\\
$G_{\rm RP}$ (mag) & 9.5732 & 0.0003 & Gaia DR3\\
$G_{\rm RVS}$ (mag) & 9.2960 & 0.0053 & Gaia DR3\\
$B$ (mag) & 11.4960 & 0.0930 & TIC\\
$V$ (mag) & 10.5110 & 0.0080 & TIC\\
$J$ (mag) & 8.7030 & 0.0290 & TIC\\
$H$ (mag) & 8.1920 & 0.0530 & TIC\\
$K$ (mag) & 8.0530 & 0.0290 & TIC\\
$W1$ (mag) & 7.9570 & 0.0240 & TIC\\
$W2$ (mag) & 8.0530 & 0.0190 & TIC\\
$W3$ (mag) & 7.9510 & 0.0200 & TIC\\
$W4$ (mag) & 7.7110 & 0.1520 & TIC\\
\hline
\enddata
\tablecomments{
Gaia DR3 magnitude uncertainties are computed from the mean flux signal-to-noise ratio as $\sigma_{\rm mag} = 2.5/(\ln 10 \times {\rm S/N})$.  Data sources are the TESS Input Catalog \citep{2019AJ....158..138S} and Gaia DR3 \citep{DR3}.
}
\end{deluxetable*}

The outer body eclipse was identified by R. Gagliano in our usual discovery process described in \citet{2021AJ....161..162P} and \citet{2025ApJS..279...50K}.  Briefly, we use a neural network to find {\em TESS} lightcurves likely containing eclipses.  We provide the candidates to our citizen scientist colleagues in the Visual Survey Group \citep[VSG;][]{2022PASP..134g4401K}, who then manually review them for the presence of extra eclipses or other interesting phenomena.  This process has contributed to many interesting discoveries, including the identification of more than 250 new quadruple star system candidates \citep{2022ApJS..259...66K,2024MNRAS.527.3995K,2026AJ....171...29K}.  Upon identification of the lightcurve of TIC 295741342, we sought follow-up spectroscopic observations.

\section{Spectroscopy}
\label{sec::spectroscopy}

The spectroscopic observations of \ticstar\ were collected at the
Center for Astrophysics, with the Tillinghast Reflector Echelle
Spectrograph \citep[TRES;][]{Furesz:2008, Szentgyorgyi:2007} on the
1.5m Tillinghast reflector at the Fred L.\ Whipple Observatory, on
Mount Hopkins (AZ, USA). The instrument delivers a resolving power of
$R \approx 44,\!000$ in 51 orders, over a wavelength range of
3800--9100~\AA. We gathered 48 spectra between 2021 October and 2026
January, with signal-to-noise ratios of 32 to 102 per resolution
element of 6.8~\kms. The reductions were carried out with a dedicated
pipeline \citep[see][]{Buchhave:2012}, and changes in the velocity
zero-point were monitored with observations of IAU standards each run.
Observations of minor planets in the solar system were then used to
place the velocities listed below on the IAU system.

\begin{deluxetable*}{rcrrrrrrr}
\tablecaption{TRES RV measurements for TIC~295741342.\label{tab:rv_data}}
\tablecolumns{8}
\tablehead{
\colhead{\#} & \colhead{BJD$-$2\,400\,000} & \colhead{$v_{\mathrm{Aa}}$} & \colhead{$v_{\mathrm{Ab}}$} & \colhead{$v_{\mathrm{B}}$} & \colhead{$\sigma_{\mathrm{Aa}}$} & \colhead{$\sigma_{\mathrm{Ab}}$} & \colhead{$\sigma_{\mathrm{B}}$} \\ [-6pt]
\colhead{} & \colhead{(d)} & \colhead{(km\,s$^{-1}$)} & \colhead{(km\,s$^{-1}$)} & \colhead{(km\,s$^{-1}$)} & \colhead{(km\,s$^{-1}$)} & \colhead{(km\,s$^{-1}$)} & \colhead{(km\,s$^{-1}$)}
}
\startdata
 1 & 59515.0339 &    6.11 &  125.72 &  30.19 &  8.38 &  7.36 & 0.10 \\
 2 & 59524.0340 &   $-$2.31 &  126.47 &  28.31 &  8.83 &  7.75 & 0.10 \\
 3 & 59537.0415 &   69.04 &   69.75 &  25.43 &  9.09 &  7.99 & 0.11 \\
 4 & 59545.9101 &  121.06 &   $-$1.10 &  23.50 &  7.00 &  6.15 & 0.09 \\
 5 & 59565.8411 &   42.62 &  104.43 &  20.06 &  7.10 &  6.24 & 0.09 \\
 6 & 59585.8823 &   $-$4.20 &  150.21 &  18.99 &  8.45 &  7.43 & 0.10 \\
 7 & 59606.7201 &  130.74 &   $-$3.60 &  23.96 &  7.26 &  6.38 & 0.09 \\
 8 & 59626.7553 &  118.63 &  $-$12.30 &  37.90 &  7.09 &  6.23 & 0.09 \\
 9 & 59665.6633 &    4.30 &   46.18 &  67.06 &  7.41 &  6.51 & 0.09 \\
10 & 59687.6417 &   97.60 &  $-$58.08 &  71.96 &  6.87 &  6.04 & 0.07 \\
11 & 59930.9321 &  127.75 &    1.56 &  29.40 &  5.90 &  5.19 & 0.06 \\
12 & 59934.8766 &  126.05 &   $-$9.85 &  28.52 &  6.75 &  5.93 & 0.07 \\
13 & 59949.8286 &  140.78 &   $-$6.27 &  25.25 &  6.48 &  5.69 & 0.07 \\
14 & 59970.8091 &   $-$3.70 &  148.18 &  21.10 &  6.57 &  5.77 & 0.07 \\
15 & 60034.6327 &  123.53 &   $-$4.78 &  33.79 &  7.14 &  6.28 & 0.09 \\
16 & 60051.6362 &  $-$29.68 &  119.58 &  49.07 &  7.10 &  6.24 & 0.09 \\
17 & 60060.6264 &  $-$12.10 &   85.54 &  56.60 &  7.24 &  6.36 & 0.09 \\
18 & 60283.9577 &   13.24 &   92.78 &  42.53 &  5.10 &  4.48 & 0.05 \\
19 & 60291.9186 &  134.69 &  $-$33.36 &  40.78 &  5.30 &  4.66 & 0.06 \\
20 & 60310.9193 &  136.39 &  $-$33.47 &  36.81 &  4.99 &  4.38 & 0.05 \\
21 & 60327.8210 &  $-$13.72 &  133.21 &  33.03 &  4.88 &  4.29 & 0.05 \\
22 & 60338.8080 &  112.17 &    6.84 &  30.58 &  4.90 &  4.30 & 0.05 \\
23 & 60352.8380 &   92.36 &   32.23 &  27.49 &  5.28 &  4.64 & 0.06 \\
24 & 60357.7800 &  113.95 &   \dots &  26.32 &  4.21 &  \dots & 0.05 \\
25 & 60363.7112 &  132.74 &   $-$1.53 &  25.03 &  5.36 &  4.71 & 0.06 \\
26 & 60372.6842 &  151.28 &  $-$24.29 &  23.21 &  5.07 &  4.45 & 0.05 \\
27 & 60391.7122 &  158.19 &  $-$21.34 &  20.00 &  5.30 &  4.66 & 0.06 \\
28 & 60412.6349 &    3.74 &  139.24 &  19.05 &  6.00 &  5.27 & 0.06 \\
29 & 60420.6675 &  147.45 &   $-$6.06 &  20.22 &  6.15 &  5.40 & 0.07 \\
30 & 60603.0144 &  $-$27.82 &  106.95 &  61.13 &  3.30 &  2.90 & 0.04 \\
31 & 60608.0149 &   \dots &  115.60 &  60.17 &  \dots &  2.65 & 0.03 \\
32 & 60622.0250 &  $-$24.75 &  114.87 &  57.55 &  3.98 &  3.50 & 0.04 \\
33 & 60629.9860 &  113.78 &  $-$28.74 &  55.93 &  3.88 &  3.41 & 0.04 \\
34 & 60654.9188 &    3.14 &   89.90 &  51.05 &  4.10 &  3.61 & 0.04 \\
35 & 60664.8885 &  $-$26.54 &  120.56 &  49.01 &  3.98 &  3.50 & 0.04 \\
36 & 60678.8591 &   $-$7.87 &  111.41 &  46.20 &  3.03 &  2.66 & 0.03 \\
37 & 60687.8617 &   \dots &   64.14 &  44.28 &  \dots &  2.83 & 0.03 \\
38 & 60693.9031 &  $-$19.40 &  129.42 &  43.05 &  3.14 &  2.76 & 0.03 \\
39 & 60707.8273 &  $-$19.83 &  130.14 &  40.15 &  3.21 &  2.83 & 0.03 \\
40 & 60712.7181 &  $-$18.92 &  131.09 &  39.02 &  2.98 &  2.61 & 0.03 \\
41 & 60714.7167 &  125.76 &  $-$17.02 &  38.58 &  3.13 &  2.74 & 0.03 \\
42 & 60724.7114 &  138.42 &  $-$29.21 &  36.42 &  3.29 &  2.89 & 0.04 \\
43 & 60979.0250 &  $-$56.39 &  108.78 &  67.46 &  3.94 &  3.47 & 0.04 \\
44 & 60986.0217 &  119.04 &  $-$61.40 &  66.33 &  2.92 &  2.56 & 0.03 \\
45 & 60993.0246 &  $-$50.38 &   \dots &  65.14 &  4.40 &  \dots & 0.05 \\
46 & 61006.9597 &  $-$18.40 &   88.04 &  62.64 &  2.87 &  2.52 & 0.03 \\
47 & 61021.9392 &   \dots &  113.95 &  59.81 &  \dots &  2.81 & 0.03 \\
48 & 61052.8876 &  125.68 &   \dots &  53.88 &  3.32 &  \dots & 0.04
\enddata
\tablecomments{Ellipses indicate epochs where a given star's RV could not be reliably measured.}
\end{deluxetable*}

Initial inspection of the spectra showed only the prominent lines of
the giant component in the system (star~B), which dominates the
light. We measured its radial velocity (RV) by cross-correlation,
using an appropriate synthetic template from a pre-computed library
based on model atmospheres by R.\ L.\ Kurucz, and a line list tuned
manually by Jon Morse to improve the match to real spectra. 
These templates cover the region of
the \ion{Mg}{1}\,b triplet ($\sim$5187~\AA), which captures most
of the velocity information.
Initial estimates of the stellar parameters for star~B were obtained using the Stellar
Parameters Classification (SPC) tool \citep{Buchhave:2012, Buchhave:2014},
ignoring the presence of the binary. This
yielded an effective temperature of $T_{\rm eff} = 4994 \pm 36$~K,
a surface gravity of $\log g = 2.810 \pm 0.063$, a metallicity estimate of
${\rm [m/H]} = -0.325 \pm 0.035$, and a projected rotational velocity of
$v \sin i = 3.59 \pm 0.22~\kms$, which includes the contribution from
macroturbulence. The uncertainties above are formal only.
Given the discreteness of
our stellar library, the selection of the optimal template was made
by running grids of cross-correlations centered on the SPC values,
as described by \cite{Torres:2002}, to identify the best match.
We settled on a template with $T_{\rm eff} = 5000$~K, $\log g = 3.5$,
${\rm [m/H]} = -0.5$, and $v \sin i = 4~\kms$. The preliminary velocities
of star B showed clear variability with a period of 412~d.

Further inspection of the spectra with TODCOR, a two-dimensional
cross-correlation technique \citep{Zucker:1994}, revealed the faint
lines of the primary of the eclipsing binary (star~Aa), and hints of
the secondary. As reported below, each of these two components contributes only
3\% or less of the total flux at the mean wavelength of our observations.
An extension of TODCOR to three dimensions
\citep[TRICOR;][]{Zucker:1995} then allowed us to measure the
velocities of all three stars more accurately. Because Aa and Ab are
too faint for us to establish their template parameters independently,
we adopted surface gravities of $\log g = 4.5$, appropriate for
main-sequence objects, and representative temperatures of 6500~K for
both stars, guided by preliminary mass estimates.
Rotational velocities were set to 12 and 10~\kms, respectively, on the
assumption that the stars' spin is approximately synchronized with the
orbital motion. Preliminary spectroscopic orbital solutions showed
that the RVs in the inner, 4.75~d orbit include a significant
contribution from the motion of the binary in the outer orbit.

The final radial velocities and uncertainties for the three components of
\ticstar\ are reported in Table \ref{tab:rv_data}. In addition to the velocities, we
used TRICOR to infer the fractional flux contributions of the stars at
the mean wavelength of our observations ($\sim$5187~\AA). They are as
follows: $F_{\rm Aa} = 0.0305 \pm 0.0061$, $F_{\rm Ab} = 0.0266 \pm
0.0089$, and $F_{\rm B} = 0.943 \pm 0.013$.

\section{Spectro-Photodynamical Model}
\label{sec::model}

Our model has been developed over the course of years of effort and thousands of simulations of a variety of multiple star systems.  Over time, we have added features and refined others, creating a rather comprehensive spectro-photodynamical model that simultaneously fits the lightcurve, RVs, eclipse times, and the spectral energy distribution (SED), and is capable of being adapted for any hierarchy. In this section, we will discuss the various model capabilities.

\subsection{MIST Isochrone Interpolation}
\label{sec::mist_model}

The MESA Isochrones \& Stellar Tracks \citep[MIST;][]{2016ApJS..222....8D,2016ApJ...823..102C,2011ApJS..192....3P,2013ApJS..208....4P,2015ApJS..220...15P} provide a means of interpolating various properties of stellar evolution models.  We downloaded the MIST isochrones\footnote{\url{https://waps.cfa.harvard.edu/MIST/interp_isos.html}} for [Fe/H] range [-4.0,0.5] and log age range [5.1,10.1], both in increments of 0.1.  

With inputs of initial mass, [Fe/H], and log age, we return interpolated values of mass, log specific gravity (log g), log bolometric luminosity (log L), log radius (log R), and log effective temperature (log T${_{\text{eff}}}$). After applying a bolometric correction for the {\em TESS} passband (600-1000 nm) from the luminosity and T${_{\text{eff}}}$, we obtained the relative luminosities of each star in the system.  The values returned from the interpolation were then used in modeling the system dynamics, the lightcurve, and the SED.

\subsection{N-body Dynamical Simulation}
\label{sec::nbody_model}

With the masses from interpolation of the MIST isochrones, we used the {\sc REBOUND} N-body simulator code \citep{Rein12} to model the dynamics of the triple star system.  We combine the stars into a simulation centered on the full system center of mass, viewed from the negative $z$ direction.  Once we build the system configuration, we integrate for the duration of the available data using the highly accurate IAS15 integrator \citep{2015MNRAS.446.1424R}, which is a 15th-order adaptive timestep integration method that reduces the timestep during close interactions to maintain the fidelity of the simulation.  The outputs from this integration are the cartesian coordinates of the stars at the times of each simulated data point, as well as the z-component velocities that were used to fit the RV data.

\subsection{Photometric Lightcurve Modeling}
\label{sec::lc_model}

After the coordinate history of all stars is known from the N-body simulation, we proceed to construct the lightcurve of the system.  For this process, we used the radii, T$_{\text{eff}}$, and bandpass-corrected bolometric luminosity from the MIST interpolation for each star. For each timestep in the model we track the positions of the stars and determine if there is an overlap between any number of them.  Pairwise eclipses are computed using the standard circular lens formula. When three stars have a combined overlap, the total occluded area is computed via inclusion-exclusion, which requires the area of the three-circle intersection to be calculated geometrically.  The per-timestep visible area of each star combined with the relative luminosities translates directly to the relative flux, allowing for the construction of the lightcurve for the duration of the simulation.  We temporally shift the eclipses as required by the Light Travel Time Effect (LTTE) described in the next section.

\subsection{Eclipse Times and ETV Modeling}
\label{sec::etv_model}

Per Equation 1 of \citet{2015MNRAS.448..946B}, ETVs for a triple star system are governed by a general form as follows:
\begin{eqnarray}
\Delta&=&T(E)-T(0)-P_\mathrm{s}E \nonumber \\
&=&\sum_{i=0}^{2}c_iE^i+\left[\Delta_\mathrm{LTTE}+\Delta_\mathrm{dyn}+\Delta_\mathrm{apse}\right]_0^E,
\end{eqnarray}
where $T(E)$ is the time of the eclipse, $T(0)$ is the reference epoch, $P_s$ is the sidereal (eclipsing) period, $E$ is the cycle number of the binary since $T(0)$, $c_i$ are fitted constants of a polynomial in $E$, $\Delta_\mathrm{dyn}$ are dynamical effects, and $\Delta_\mathrm{apse}$ are apsidal motion effects.

The numerical integration of the N-body simulation accounts for the dynamical and apsidal effects, but does not consider the LTTE, which is the classical Roemer delay as given by Equation 2 of \citet{2015MNRAS.448..946B}:
\begin{equation}
\Delta_\mathrm{LTTE}=-\frac{a_\text{out}\sin i_\text{out}}{c}\frac{\left(1-e_\text{out}^2\right)\sin(v_\text{out}+\omega_\text{out})}{1+e_\text{out}\cos v_\text{out}},
\label{eqn:ltte} 
\end{equation}
where, for the outer orbit, $a$ is the semi-major axis, $i$ is the inclination, $e$ is the eccentricity, $v$ is the true anomaly, and $\omega$ is the argument of periastron. After modeling the light curve, we temporally shift it by $\Delta_\mathrm{LTTE}$.

To measure the eclipse times and obtain the corresponding ETVs, we followed the methodology of \citet{2022ApJS..259...66K}. Briefly, we start with a standard procedure of utilizing BLS to measure a preliminary orbital period and ephemeris. Next, we combine these with a generalized Gaussian function (Equation 1 of \citealt{2022ApJS..259...66K}) that provides a flexible yet rather simple mathematical model that naturally accounts for a variety of eclipse shapes, as well as potential slight variations across sectors. We apply the function to each observed eclipse and measure the corresponding times, depths, and durations. Finally, the eclipse times are compared to a linear ephemeris and the corresponding deviations recorded as ETVs.  We provide the eclipse times and ETVs for TIC 295741342 in Table \ref{tab:etv_data}.

\begin{deluxetable*}{rc@{\hskip 6pt}r@{\hskip 6pt}r|rc@{\hskip 6pt}r@{\hskip 6pt}r}
\tablewidth{0pt}
\tablecaption{Eclipse timing measurements for TIC~295741342.\label{tab:etv_data}}
\tablehead{
\multicolumn{4}{c|}{Primary Eclipses} & \multicolumn{4}{c}{Secondary Eclipses} \\
\cline{1-4} \cline{5-8}
\colhead{Cycle} & \colhead{$T_0$} & \colhead{ETV} & \multicolumn{1}{c|}{$\sigma$} &  \colhead{Cycle} & \colhead{$T_0$} & \colhead{ETV} & \colhead{$\sigma$} \\[-6pt]
& \colhead{(BJD$-$2\,400\,000)} & \colhead{(min)} & \multicolumn{1}{c|}{(min)} &  & \colhead{(BJD$-$2\,400\,000)} & \colhead{(min)} & \colhead{(min)}
}
\startdata
  0.0 & 58495.571376 & $-$5.090 & 1.831 &  $-$0.5 & 58493.378313 & $-$1.864 & 1.992 \\
  1.0 & 58500.327161 & $-$4.188 & 2.226 &    0.5 & 58498.130713 & $-$5.837 & 1.092 \\
  3.0 & 58509.838238 & $-$3.095 & 1.768 &    1.5 & 58502.885679 & $-$6.117 & 2.781 \\
  4.0 & 58514.591726 & $-$5.501 & 2.494 &    2.5 & 58507.639763 & $-$7.665 & 3.204 \\
149.0 & 59204.091669 & $-$2.770 & 1.365 &    3.5 & 58512.396153 & $-$5.894 & 2.260 \\
150.0 & 59208.849159 &    0.586 & 1.299 &  148.5 & 59201.897148 & $-$1.828 & 1.646 \\
152.0 & 59218.362137 &    4.417 & 3.970 &  149.5 & 59206.652571 & $-$1.448 & 1.423 \\
155.0 & 59232.627358 &    4.049 & 1.244 &  150.5 & 59211.408906 &    0.243 & 1.627 \\
156.0 & 59237.383904 &    6.046 & 1.142 &  151.5 & 59216.164650 &    1.085 & 4.077 \\
158.0 & 59246.895410 &    7.757 & 2.818 &  152.5 & 59220.924217 &    7.431 & 4.800 \\
159.0 & 59251.649998 &    6.935 & 1.595 &  154.5 & 59230.430950 &    2.266 & 1.302 \\
457.0 & 60668.683552 &    1.446 & 1.085 &  155.5 & 59235.189356 &    6.941 & 1.619 \\
458.0 & 60673.435494 & $-$3.186 & 0.932 &  156.5 & 59239.944567 &    7.016 & 1.844 \\
459.0 & 60678.192801 & $-$0.093 & 1.010 &  157.5 & 59244.702363 &   10.812 & 1.393 \\
460.0 & 60682.946183 & $-$2.652 & 1.412 &  158.5 & 59249.456201 &    8.908 & 1.235 \\
461.0 & 60687.699945 & $-$4.663 & 1.353 &  456.5 & 60666.485485 & $-$3.096 & 1.345 \\
 \dots & \dots & \dots & \dots &  457.5 & 60671.241018 & $-$2.558 & 1.150 \\
 \dots & \dots & \dots & \dots &  458.5 & 60675.994283 & $-$5.287 & 1.542 \\
 \dots & \dots & \dots & \dots &  459.5 & 60680.751854 & $-$1.815 & 1.232 \\
 \dots & \dots & \dots & \dots &  460.5 & 60685.507376 & $-$1.294 & 1.123
\enddata
\tablecomments{The ETV is defined as $\mathrm{ETV} = T_{0,\mathrm{obs}} - T_{0,\mathrm{proj}}$, where $T_{0,\mathrm{proj}}$ is the time predicted by a linear ephemeris. Primary eclipses have integer cycle numbers; secondary eclipses have half-integer cycle numbers. Cycle numbers are computed relative to an epoch of BJD~2\,458\,495.571376 with period $P = 4.7558$~d.}
\end{deluxetable*}

\subsection{SED Modeling}
\label{sec::sed_model}
In order to model the SED of each of the stars in the system, we used synthetic BT-Settl model atmospheres \citep{2014IAUS..299..271A}\footnote{Available from \url{http://svo2.cab.inta-csic.es/theory/newov2/}.} to build an interpolator over the T$_{\text{eff}}$ range 2,000 - 30,000 K, returning the energy per unit wavelength for the range 1-30,000 nm at the star.  We utilize the T$_{\text{eff}}$ from the MIST isochrone interpolation to find the SED from the BT-Settl atmospheres and, to model the energy received at Earth, we multiply the SED at the star by the square the radius of the star divided by the distance to the star. We use the interstellar extinction, $A_v$, as a free parameter, modeling the effect of $A_v$ on the SED per equations 1-5 of \citet{1989ApJ...345..245C} and calculating the $\chi^2$ of the SED fit at the star.

\subsection{Solid Angle Loss}
\label{sec::add_error_model}

With such a large parameter space, we found that helping to ``nudge'' the model toward a viable system configuration avoided local minima traps.  We accomplished this through an additional error term in the model $\chi^{2}$ encouraging the binary to be behind the tertiary at the time of the outer eclipse.  We call this term the ``solid angle loss'' ($L_\Omega$) and measure it only during the time of the outer eclipse.  We define $L_\Omega$ as the solid angle of the spherical cap formed by the angle between the observer's line of sight and the displacement vector from the tertiary to the binary center of mass. When the tertiary is directly in front of the binary center of mass from the observer's perspective, $L_\Omega=0$, and when the tertiary is directly behind the binary center of mass, $L_\Omega=4\pi$.  The solid angle loss smoothly decreases as the system approaches the proper configuration, making it a powerful addition to the model $\chi^2$. We apply a delta to this term, such that when $L_\Omega$ is less than a certain value, it becomes zero.  In this manner, we ``line up'' the system in the proper configuration, then let the physical measurements (lightcurve, SED, RV, and eclipse times) control the fit from that point forward.  In effect, this allows us to avoid the computationally expensive initial exploration of the huge parameter space without any influence on the final $\chi^2$ value.  So, our model $\chi^2$ is then

\begin{equation}
\chi^2_{\text{model}} = 
\begin{cases}
\chi^2_{\text{LC}} + \chi^2_{\text{SED}} + \chi^2_{\text{RV}} + \chi^2_{\text{ET}}+
\psi L_\Omega, \\ 
\text{        if } L_\Omega \geq p. \\
\chi^2_{\text{LC}} + \chi^2_{\text{SED}} + 
\chi^2_{\text{RV}} +
\chi^2_{\text{ET}}, \\ 
\text{       else.} 
\end{cases}
\end{equation}
where $p$ is defined by the user as the solid angle difference in steradians between the tertiary and the binary center of mass less than which $L_\Omega$ is no longer considered, and $\psi$ is a user-defined constant to magnify the effect of $L_\Omega$.  In our case, we used $p=0.1$ and $\psi=10^8$.  We found that using the solid angle loss for the outer orbit reduced convergence time of the model by more than half.

\subsection{Full System Model}

Based on the properties described in the preceding paragraphs, our system model has 19 free parameters:

\begin{itemize}
\item Stellar parameters:  Mass of each star. (3)
\item Orbital parameters: For the inner and outer orbits: inclination ($i$), eccentricity ($e$), argument of periastron ($\omega$), phase at time of pericenter passage ($\phi$), and the orbital period ($P$).  The time of pericenter passage ($\tau$) for each orbit is calculated as the system epoch added to $\frac{\phi P}{2\pi}$. We fix the longitude of the ascending node of the binary ($\Omega_A$) at 0$^{\circ}$, while the longitude of the ascending node of the outer orbit ($\Omega_{AB}$) is a free parameter.  (11)
\item For the full system:  Systemic velocity ($\gamma$), log age, [Fe/H], A$_v$, and parallax. (5)
\end{itemize}

We used the {\sc emcee} Markov-chain Monte Carlo (MCMC) Python implementation \citep{2013PASP..125..306F} combined with the {\sc scipy} \citep{scipy} implementation of the Differential Evolution (DE) algorithm \citep{Storn1997} to minimize the model $\chi^2$.  We start the model with a fixed number of DE iterations to narrow the parameter solution space, then use the DE population to instantiate the MCMC walkers, running the MCMC until convergence.

\section{Results}
\label{sec::results}

We applied our model to the {\em TESS} lightcurve, TRES RVs given in Table \ref{tab:rv_data}, eclipse times given in Table \ref{tab:etv_data}, and photometric fluxes from various observatories given in Table \ref{tab:stellar_params}.  Using a 64-core, 128-thread AMD EPYC 7003 ``Milan'' processor on the NASA Center for Climate Simulation (NCCS) {\em Discover} supercomputer\footnote{\url{https://www.nccs.nasa.gov/systems/discover}} with 2000 initial DE iterations and full parallelization for both DE and MCMC, our model converged to a solution in about two days.  We used 1,216 walkers (64 $\times$ 19 free parameters) each run for 50,000 steps to ensure full exploration of the parameter space.  We discarded the first 80\% of the steps as burn-in, leaving $\sim$1.2$\times10^7$ posterior samples, from which we randomly drew 100,000 to generate the values and uncertainties for the solution.

We found two solutions to the system that produced nearly identical fits but are separated completely by evolutionary state.  In one state, star B is ascending the Red Giant Branch (RGB).  In the other state, star B has completed its RGB evolution and is on the Horizontal Branch (HB).  Hereafter, we refer to these states as RGB-s and HB-s, respectively.  In RGB-s, the stars are all slightly less massive and the system is slightly older.  RGB-s parameters are given in Table \ref{tab:rgb_mcmc_results} and HB-s parameters are given in Table \ref{tab:hb_mcmc_results}.  We will first show the remarkably similar fits for each solution in this section, then discuss the substantial differences in the evolutionary states in Section \ref{sec::discussion}.

\begin{table*}
\centering
\renewcommand{\arraystretch}{1.3}
\begin{tabular}{lccc}
\toprule
\multicolumn{4}{c}{\textbf{Orbital Elements (RGB)}}\\
\midrule
 & A & AB & \\
\midrule
Period [days] & $4.7546450^{+0.0000058}_{-0.0000060}$ & $412.786^{+0.055}_{-0.055}$ \\
Semimajor axis [$R_{\odot}$] & $15.286^{+0.031}_{-0.029}$ & $364.7^{+1.2}_{-1.1}$ \\
Eccentricity & $0.06184^{+0.00033}_{-0.00027}$ & $0.3634^{+0.0011}_{-0.0011}$ \\
$\omega$ [deg] & $168.8^{+2.3}_{-2.3}$ & $260.66^{+0.22}_{-0.21}$ \\
Inclination [deg] & $88.175^{+0.040}_{-0.040}$ & $87.950^{+0.021}_{-0.022}$ \\
$\Omega$ [deg] & 0.0 (fixed) & $359.762^{+0.098}_{-0.100}$ & \\
$\tau$ [BJD] & $58489.535^{+0.092}_{-0.095}$ & $58516.6^{+7.8}_{-7.8}$ \\
Mutual Inclination [deg] & -- & $0.331^{+0.074}_{-0.063}$ & \\
\midrule
\multicolumn{4}{c}{\textbf{Stellar Parameters (RGB)}}\\
\midrule
Parameter & Aa & Ab & B \\
\midrule
\textit{m} [$M_{\odot}$] & $1.0768^{+0.0068}_{-0.0063}$ & $1.0431^{+0.0064}_{-0.0058}$ & $1.700^{+0.025}_{-0.023}$ \\
\textit{R} [$R_{\odot}$] & $1.0608^{+0.0063}_{-0.0057}$ & $1.0123^{+0.0060}_{-0.0054}$ & $10.612^{+0.096}_{-0.091}$ \\
$L_{\text{bolo}}$ [$L_{\odot}$] & $1.769^{+0.049}_{-0.044}$ & $1.501^{+0.039}_{-0.036}$ & $55.6^{+1.2}_{-1.1}$ \\
$T_{\text{eff}}$ [K] & $6464^{+35}_{-34}$ & $6351^{+33}_{-32}$ & $4839^{+13}_{-13}$ \\
Log \textit{g} [dex] & $4.4186^{+0.0023}_{-0.0025}$ & $4.4455^{+0.0024}_{-0.0026}$ & $2.6165^{+0.0052}_{-0.0052}$ \\
\midrule
\multicolumn{4}{c}{\textbf{Relative Quantities (RGB)}}\\
\midrule
Fractional Flux [in \textit{TESS}-band] & $0.02690^{+0.00013}_{-0.00012}$ & $0.023147^{+0.000052}_{-0.000053}$ & $0.94995^{+0.00012}_{-0.00013}$ \\
Fractional radius $[R/a]$ & $0.06940^{+0.00027}_{-0.00025}$ & $0.06623^{+0.00026}_{-0.00024}$ & $0.02910^{+0.00020}_{-0.00019}$ \\
\midrule
\multicolumn{4}{c}{\textbf{Global Parameters (RGB)}}\\
\midrule
Log(age) [dex] & \multicolumn{3}{c}{$9.165^{+0.017}_{-0.017}$} \\
$[\mathrm{Fe/H}]$ [dex] & \multicolumn{3}{c}{$-0.337^{+0.031}_{-0.030}$} \\
$\gamma$ [km/s] & \multicolumn{3}{c}{$46.991^{+0.026}_{-0.026}$} \\
$A_V$ [mag] & \multicolumn{3}{c}{$0.163^{+0.058}_{-0.058}$} \\
Parallax [mas] & \multicolumn{3}{c}{$1.167^{+0.024}_{-0.025}$} \\
Epoch [BJD] & \multicolumn{3}{c}{$2458491.6822742$} \\
\bottomrule
\end{tabular}
\caption{RGB-s MCMC results for the triple star system TIC~295741342. Uncertainties represent the 16th and 84th percentiles of the posterior distribution. Stellar parameters interpolated from \texttt{MIST} isochrones and uncertainties do not account for possible systematics in the stellar models. Orbital element $\omega$ is reported as the secondary relative to the primary.}
\label{tab:rgb_mcmc_results}
\end{table*}
\begin{table*}
\centering
\renewcommand{\arraystretch}{1.3}
\begin{tabular}{lccc}
\toprule
\multicolumn{4}{c}{\textbf{Orbital Elements (HB)}}\\
\midrule
 & A & AB & \\
\midrule
Period [days] & $4.7546406^{+0.0000062}_{-0.0000060}$ & $412.776^{+0.055}_{-0.055}$ \\
Semimajor axis [$R_{\odot}$] & $15.424^{+0.027}_{-0.024}$ & $369.9^{+1.0}_{-1.0}$ \\
Eccentricity & $0.06177^{+0.00032}_{-0.00025}$ & $0.3619^{+0.0011}_{-0.0011}$ \\
$\omega$ [deg] & $169.0^{+2.3}_{-2.2}$ & $260.55^{+0.21}_{-0.21}$ \\
Inclination [deg] & $88.086^{+0.040}_{-0.040}$ & $87.995^{+0.058}_{-0.027}$ \\
$\Omega$ [deg] & 0.0 (fixed) & $359.765^{+0.098}_{-0.100}$ & \\
$\tau$ [BJD] & $58489.484^{+0.097}_{-0.096}$ & $58515.1^{+7.8}_{-7.8}$ \\
Mutual Inclination [deg] & -- & $0.249^{+0.092}_{-0.087}$ & \\
\midrule
\multicolumn{4}{c}{\textbf{Stellar Parameters (HB)}}\\
\midrule
Parameter & Aa & Ab & B \\
\midrule
\textit{m} [$M_{\odot}$] & $1.1061^{+0.0060}_{-0.0053}$ & $1.0719^{+0.0055}_{-0.0048}$ & $1.806^{+0.022}_{-0.024}$ \\
\textit{R} [$R_{\odot}$] & $1.0934^{+0.0061}_{-0.0051}$ & $1.0451^{+0.0056}_{-0.0048}$ & $10.64^{+0.10}_{-0.23}$ \\
$L_{\text{bolo}}$ [$L_{\odot}$] & $2.043^{+0.050}_{-0.122}$ & $1.737^{+0.042}_{-0.097}$ & $63.4^{+1.3}_{-2.9}$ \\
$T_{\text{eff}}$ [K] & $6590^{+42}_{-83}$ & $6474^{+39}_{-77}$ & $4991^{+13}_{-13}$ \\
Log \textit{g} [dex] & $4.4040^{+0.0022}_{-0.0025}$ & $4.4297^{+0.0022}_{-0.0025}$ & $2.6426^{+0.0119}_{-0.0075}$ \\
\midrule
\multicolumn{4}{c}{\textbf{Relative Quantities (HB)}}\\
\midrule
Fractional Flux [in \textit{TESS}-band] & $0.02679^{+0.00014}_{-0.00013}$ & $0.023145^{+0.000052}_{-0.000053}$ & $0.95007^{+0.00012}_{-0.00013}$ \\
Fractional radius $[R/a]$ & $0.07089^{+0.00027}_{-0.00023}$ & $0.06776^{+0.00025}_{-0.00022}$ & $0.02876^{+0.00023}_{-0.00053}$ \\
\midrule
\multicolumn{4}{c}{\textbf{Global Parameters (HB)}}\\
\midrule
Log(age) [dex] & \multicolumn{3}{c}{$9.0931^{+0.0375}_{-0.0059}$} \\
$[\mathrm{Fe/H}]$ [dex] & \multicolumn{3}{c}{$-0.353^{+0.050}_{-0.043}$} \\
$\gamma$ [km/s] & \multicolumn{3}{c}{$46.984^{+0.026}_{-0.026}$} \\
$A_V$ [mag] & \multicolumn{3}{c}{$0.351^{+0.058}_{-0.058}$} \\
Parallax [mas] & \multicolumn{3}{c}{$1.159^{+0.025}_{-0.025}$} \\
Epoch [BJD] & \multicolumn{3}{c}{$2458491.6822742$} \\
\bottomrule
\end{tabular}
\caption{HB-s MCMC results for the triple star system TIC~295741342. Uncertainties represent the 16th and 84th percentiles of the posterior distribution. Stellar parameters interpolated from \texttt{MIST} isochrones and uncertainties do not account for possible systematics in the stellar models. Orbital element $\omega$ is reported as the secondary relative to the primary.}
\label{tab:hb_mcmc_results}
\end{table*}

\begin{figure*}
   \centering
    \includegraphics[width=0.49\linewidth]{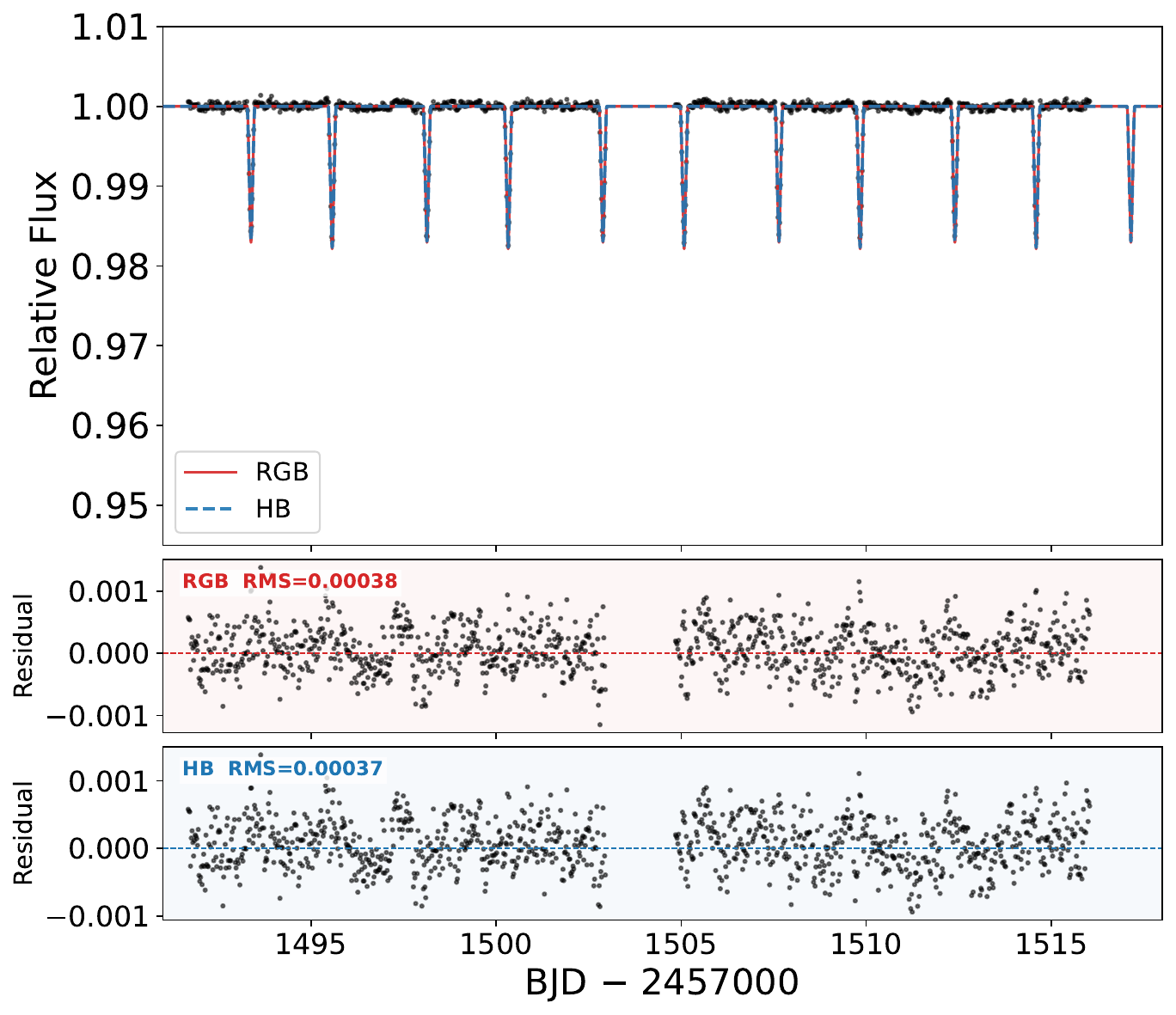}
    \includegraphics[width=0.49\linewidth]{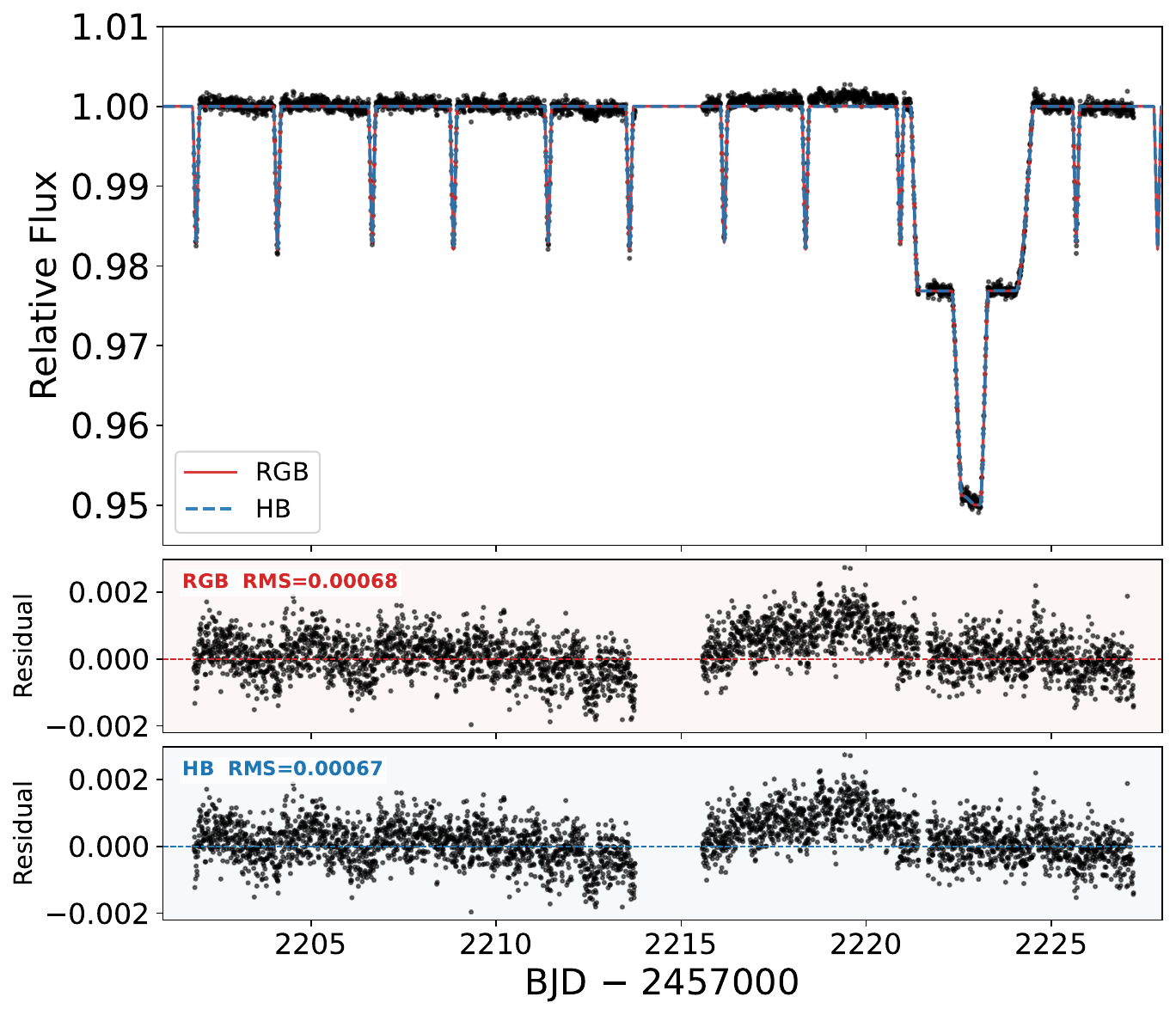}
    \includegraphics[width=0.49\linewidth]{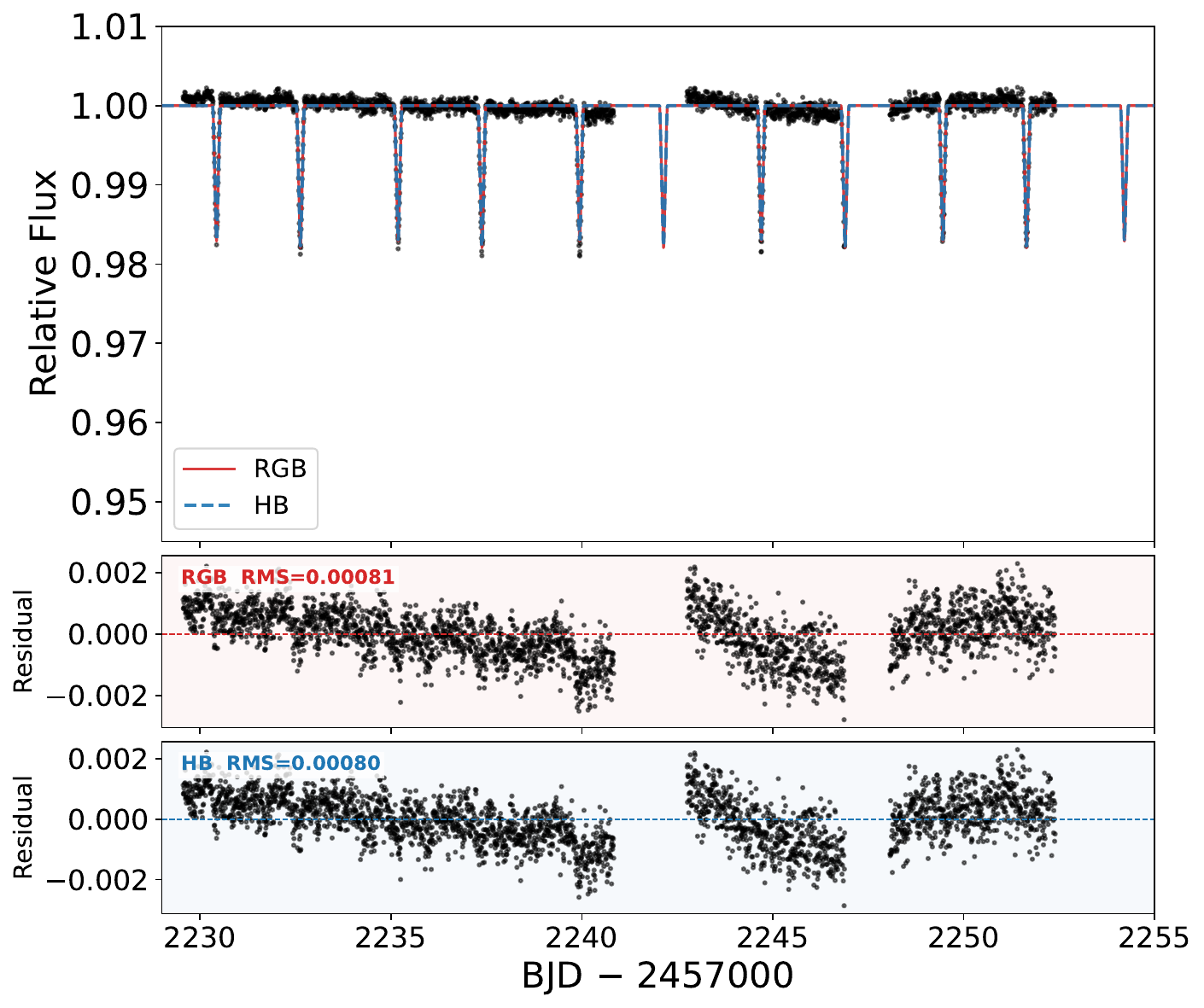}
    \includegraphics[width=0.49\linewidth]{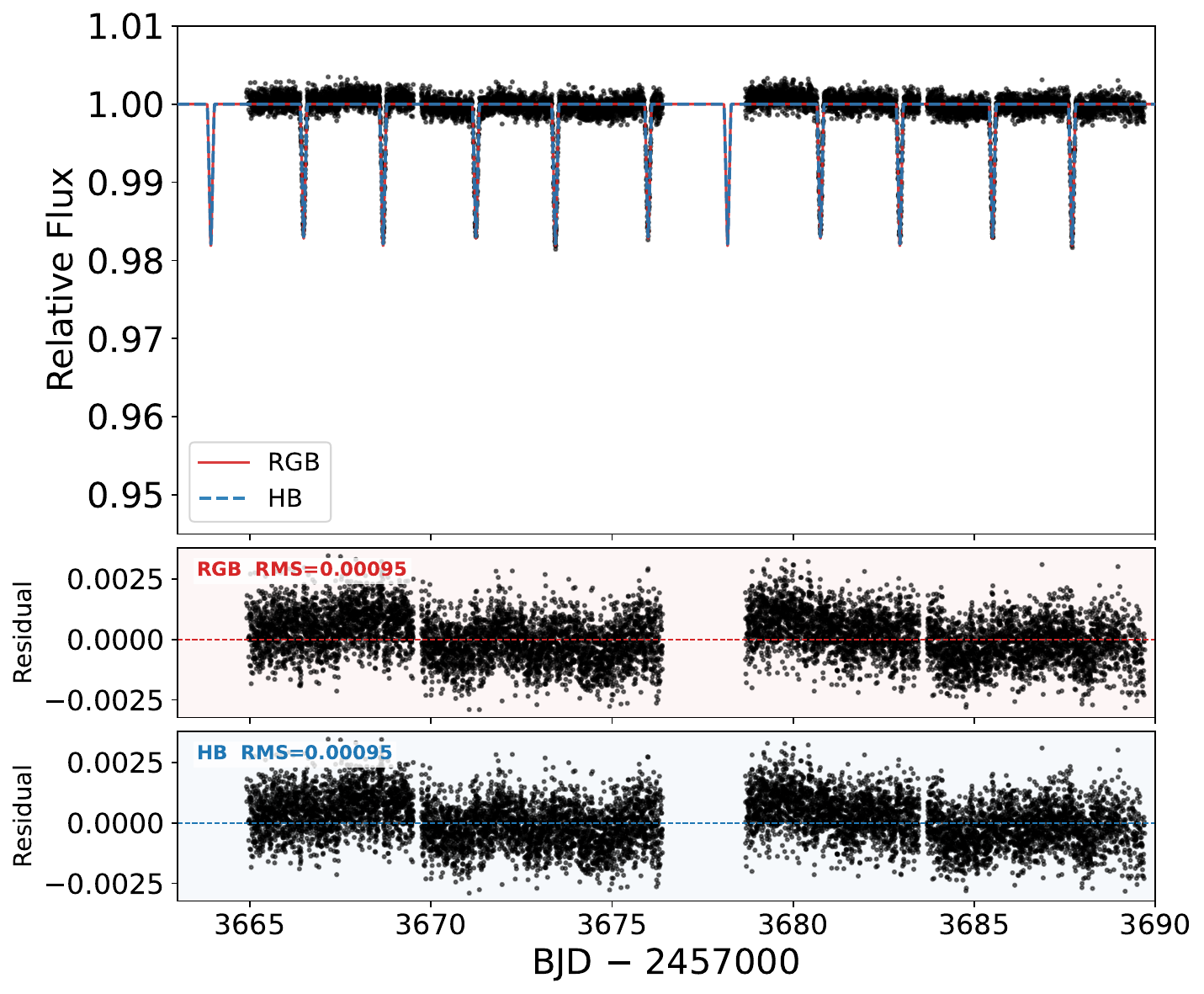}
    \caption{{\em TESS} Full-Frame Image QLP lightcurves (black) for TIC 295741342 from Sectors 7 ({\em upper left}), 33 ({\em upper right}), 34 ({\em lower left}), and 87 ({\em lower right}), plotted against the RGB-s model fit (solid red line) and the HB-s model fit (dashed blue line) -- the two fits are nearly identical so they are overlapping in the plot.  Differences can be seen in the residuals.  Residuals are shown in panels below each sector for RGB-s (shaded in red) and HB-s (shaded in blue).}
    \label{fig:lcfit}
\end{figure*}

The model fits to the lightcurve are shown in Figure \ref{fig:lcfit}, with the RGB-s fit as a solid red line and the HB-s fit as a dashed blue line.  The lightcurve data points are in black.  Sector 7 is shown in the upper left panel, Sector 33 in the upper right panel, Sector 34 in the lower left panel, and Sector 87 in the lower right panel, with separate residual panels for RGB-s (shaded in red) and the HB-s (shaded in blue).  The changing density in lightcurve data points is explained by the {\em TESS} 30-minute cadence in Sector 7, 10-minute cadence in Sectors 33 and 34, and 200-second cadence in Sector 87.  In order to ensure equal consideration of all sectors during fitting, we downsampled the fitted data to a 30-minute cadence over all sectors.  The residuals beneath each panel demonstrate an excellent fit for both solutions to the eclipsing binary as well as the Sector 33 outer eclipse.

\begin{figure*}
   \centering
    \includegraphics[width=0.48\linewidth]{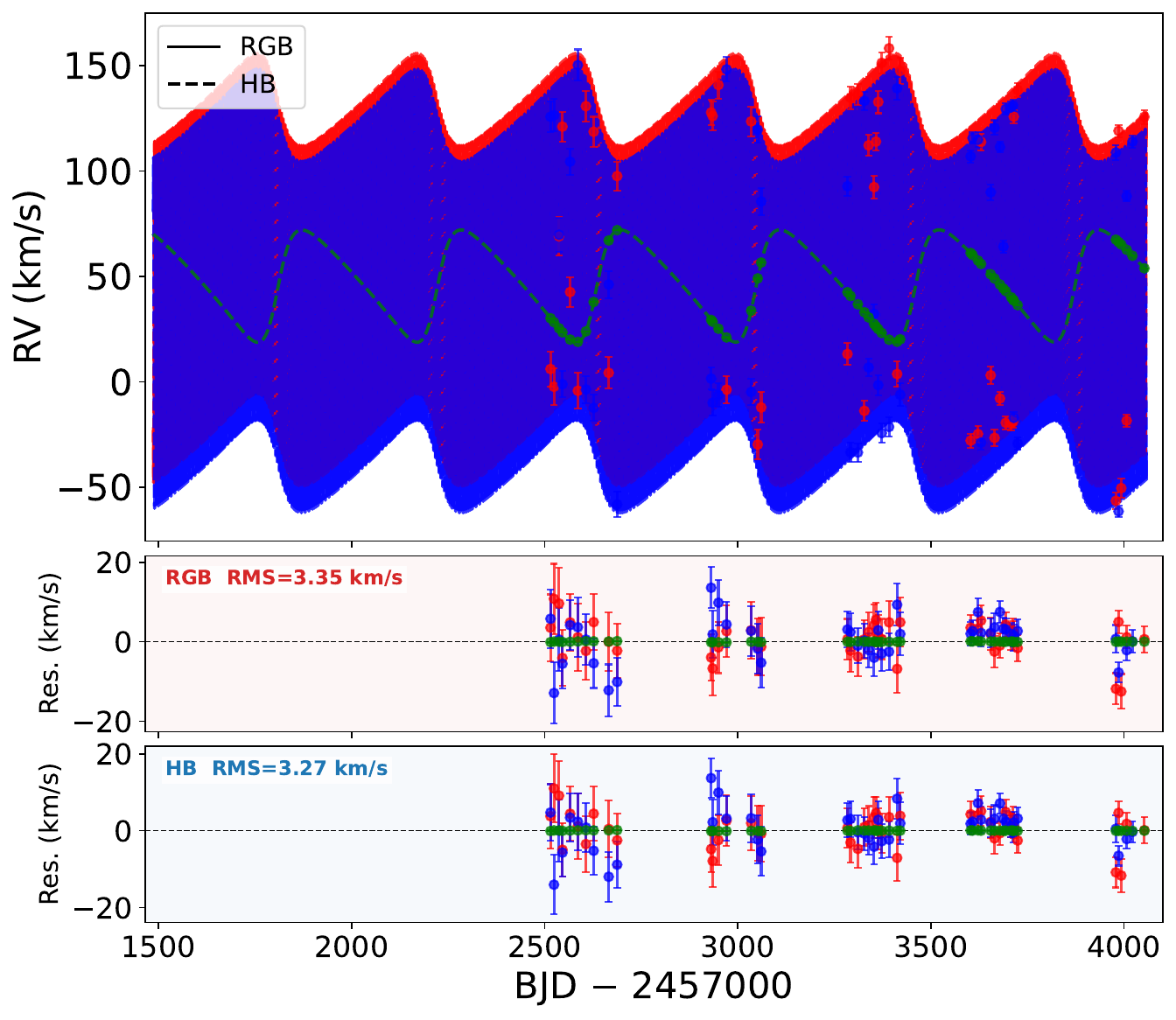}
    \includegraphics[width=0.48\linewidth]{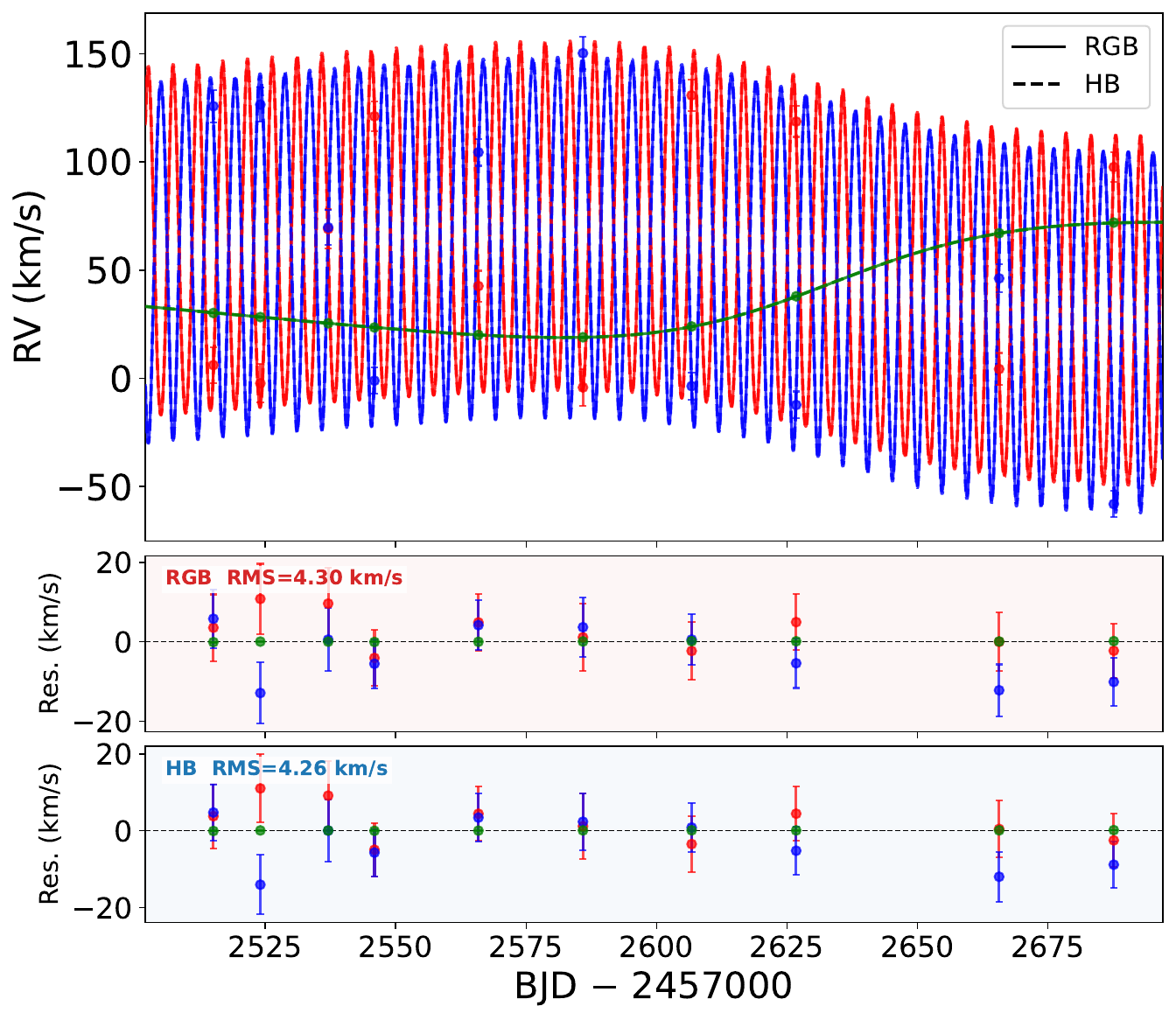}
    \includegraphics[width=0.48\linewidth]{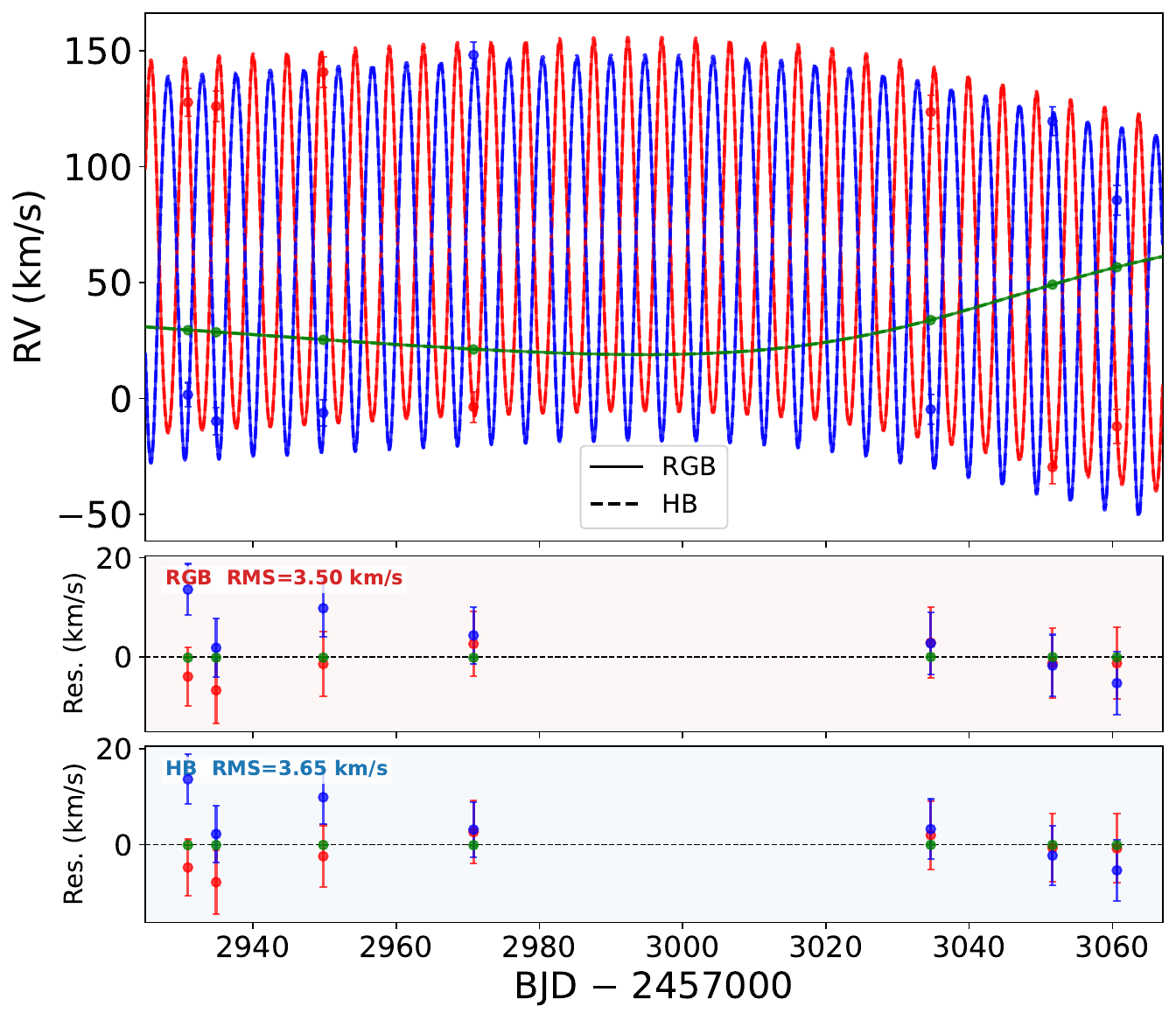}
    \includegraphics[width=0.48\linewidth]{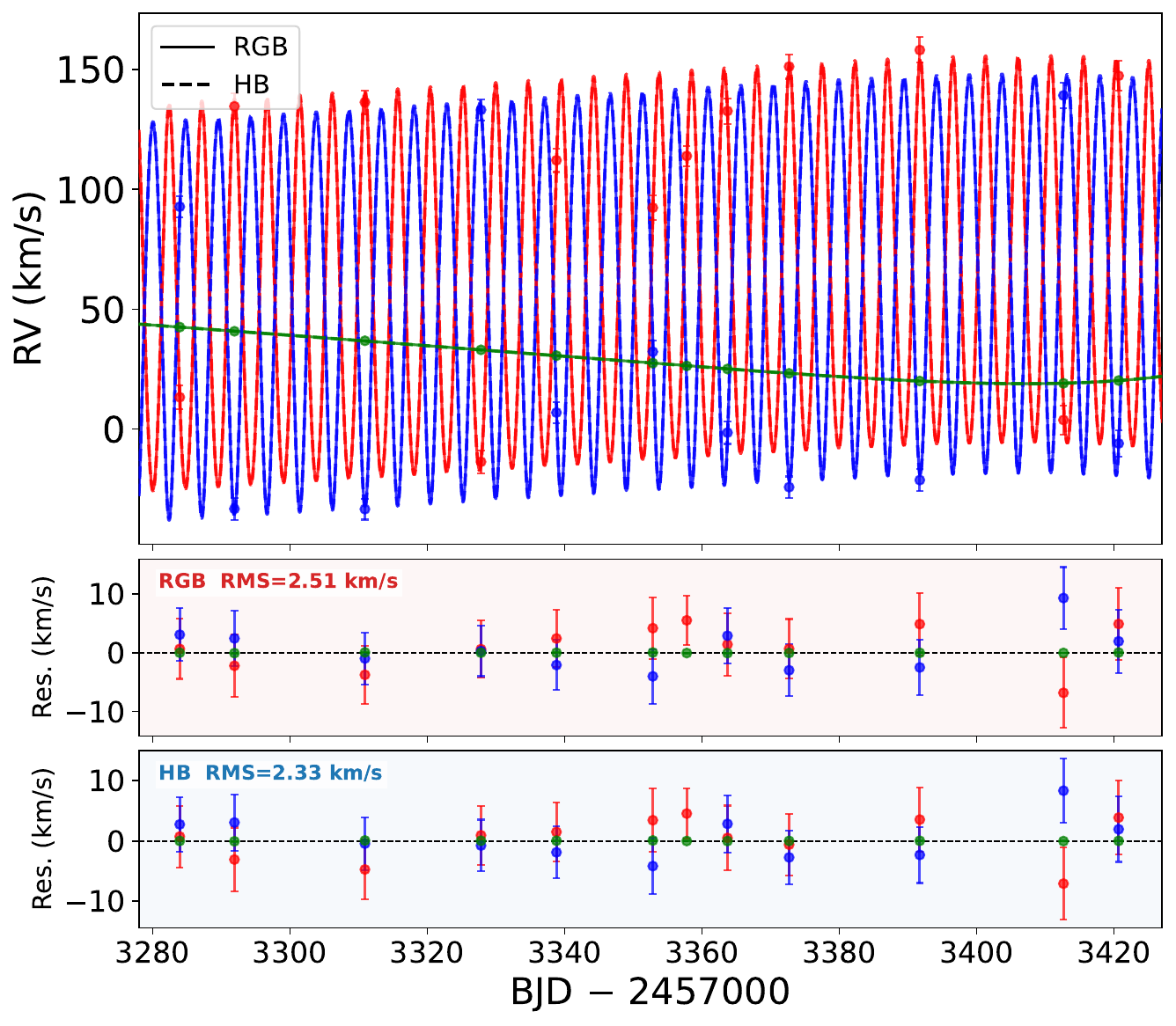}
    \includegraphics[width=0.48\linewidth]{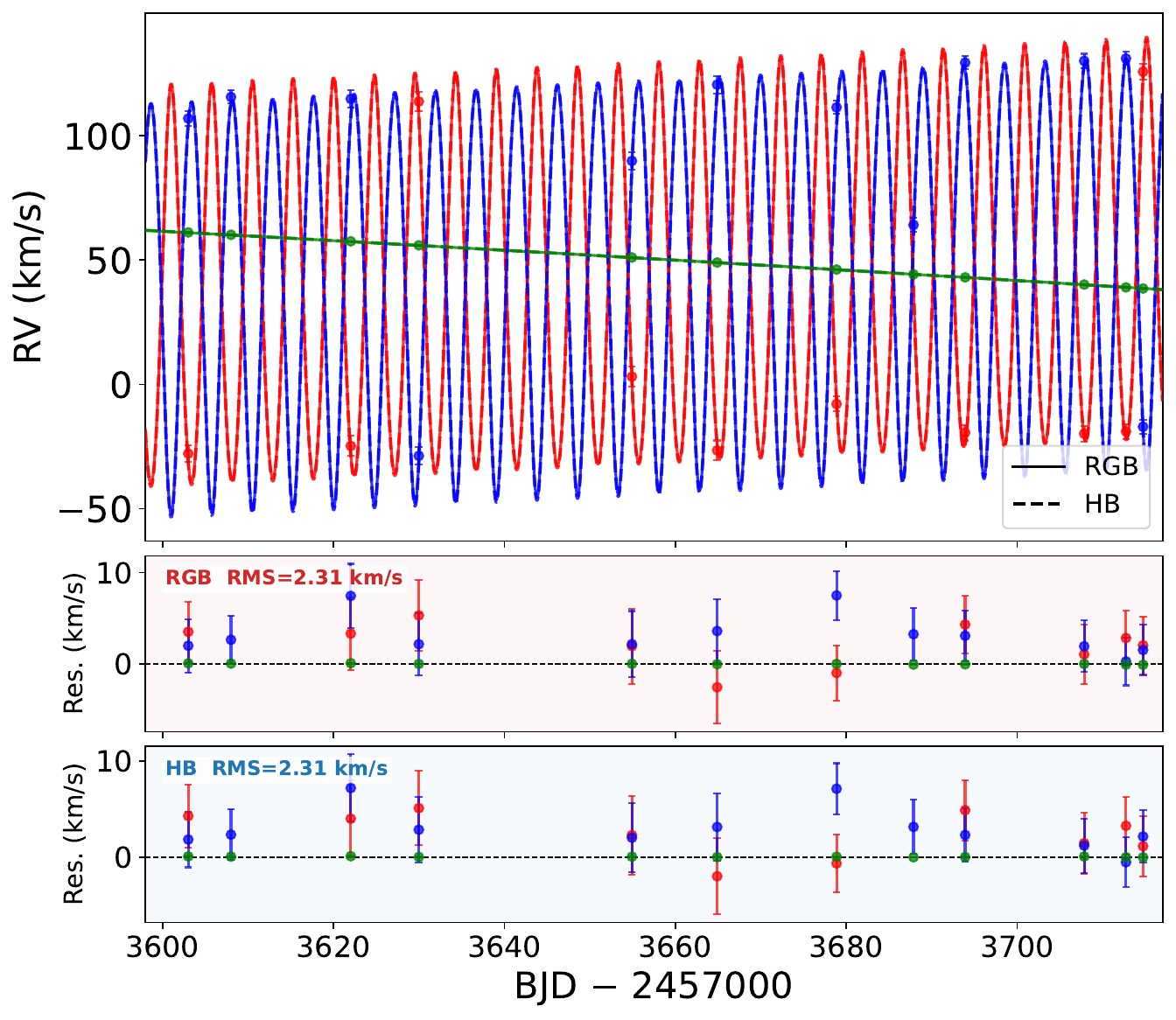}
    \includegraphics[width=0.48\linewidth]{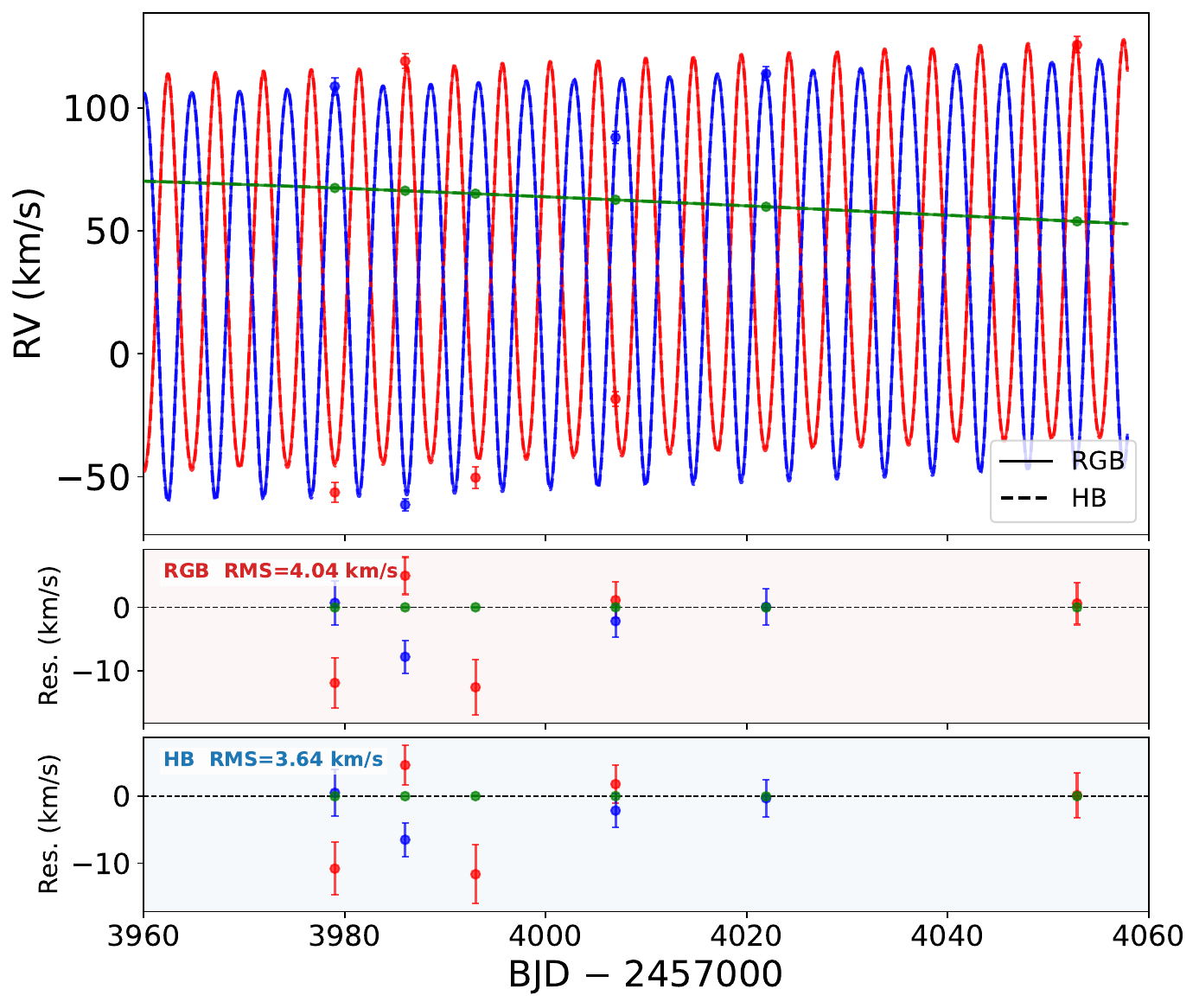}
    \caption{TRES RV measurements (points) and model RVs (lines) for stars Aa (red), Ab (blue), and B (green) for the duration of the model simulation ({\em top left}) and at time segments BJD 2459502 - 2459697 ({\em top right}), BJD 2459925 - 2460067 ({\em middle left}), BJD 2460278 - 2460427 ({\em middle right}), BJD 2460598 - 2460717 ({\em bottom left}), and BJD 2460960 - 2461060 ({\em bottom right}).  Each of the five panels (after the first panel) represents a zoom in on all of the RV points from a single observing season. The RGB-s fit is shown as solid lines and the HB-s fit as dashed lines.  As with the lightcurve fit, they are nearly identical and overlap in the plots.  Differences can be seen in the residuals.}
    \label{fig:allrv}
\end{figure*}

The overall model fit to the TRES RVs is shown in the upper left panel of Figure \ref{fig:allrv}, with star Aa in red, star Ab in blue, and star B in green.  RGB-s is shown with solid lines, while HB-s is shown with dashed lines.  The other panels show a zoomed-in view of the model fit from five separate time segments of dedicated TRES spectra collection.  The residuals are shown beneath each panel, again with the RGB-s residuals shaded in red and the HB-s residuals shaded in blue.  TRES RVs and errors correspond to Table \ref{tab:rv_data}.  The TRES measurements were planned for thorough coverage of the outer orbital phases.

\begin{figure}
    \centering
    \includegraphics[width=1.\columnwidth]{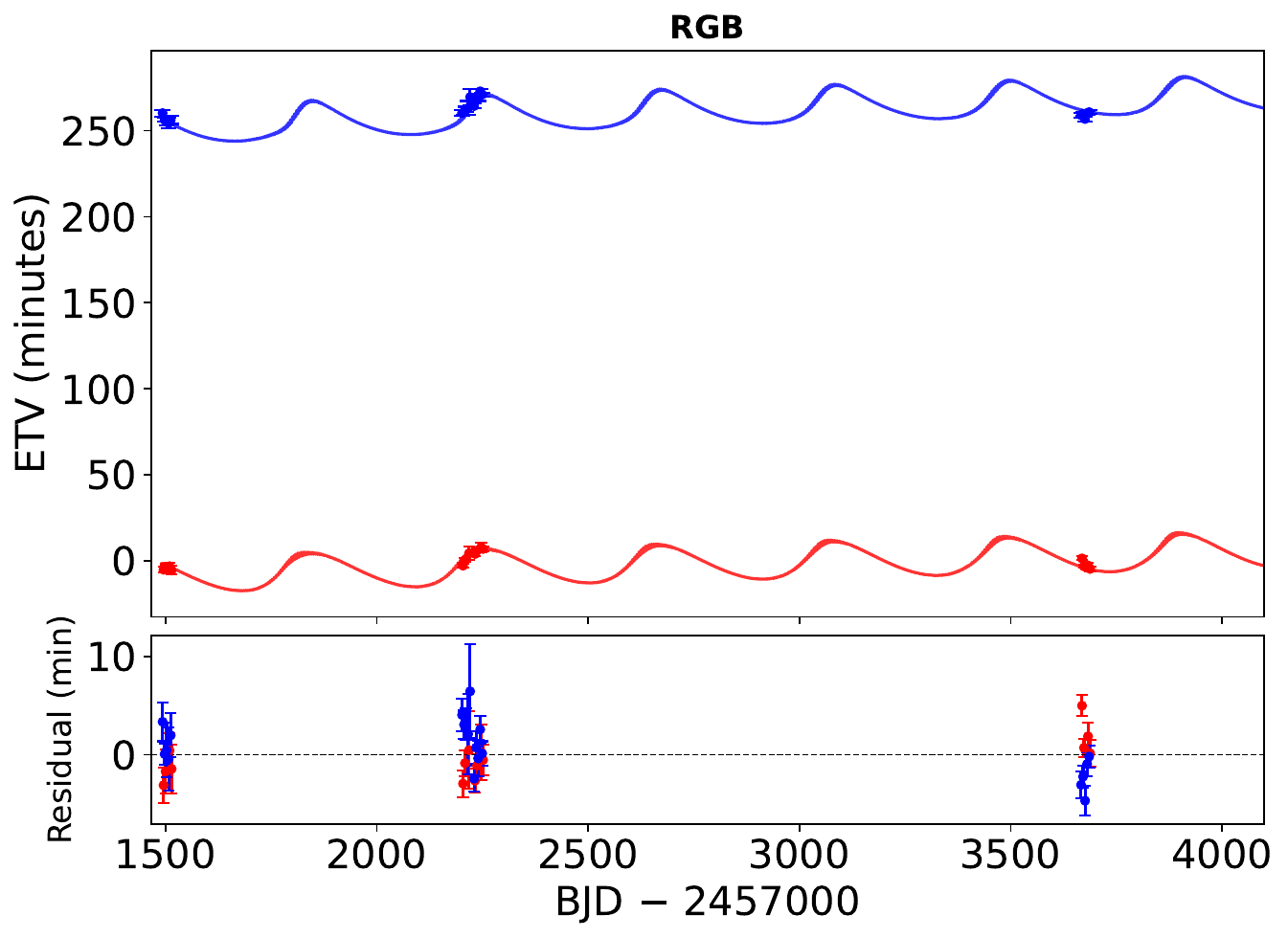}
    \includegraphics[width=1.\columnwidth]{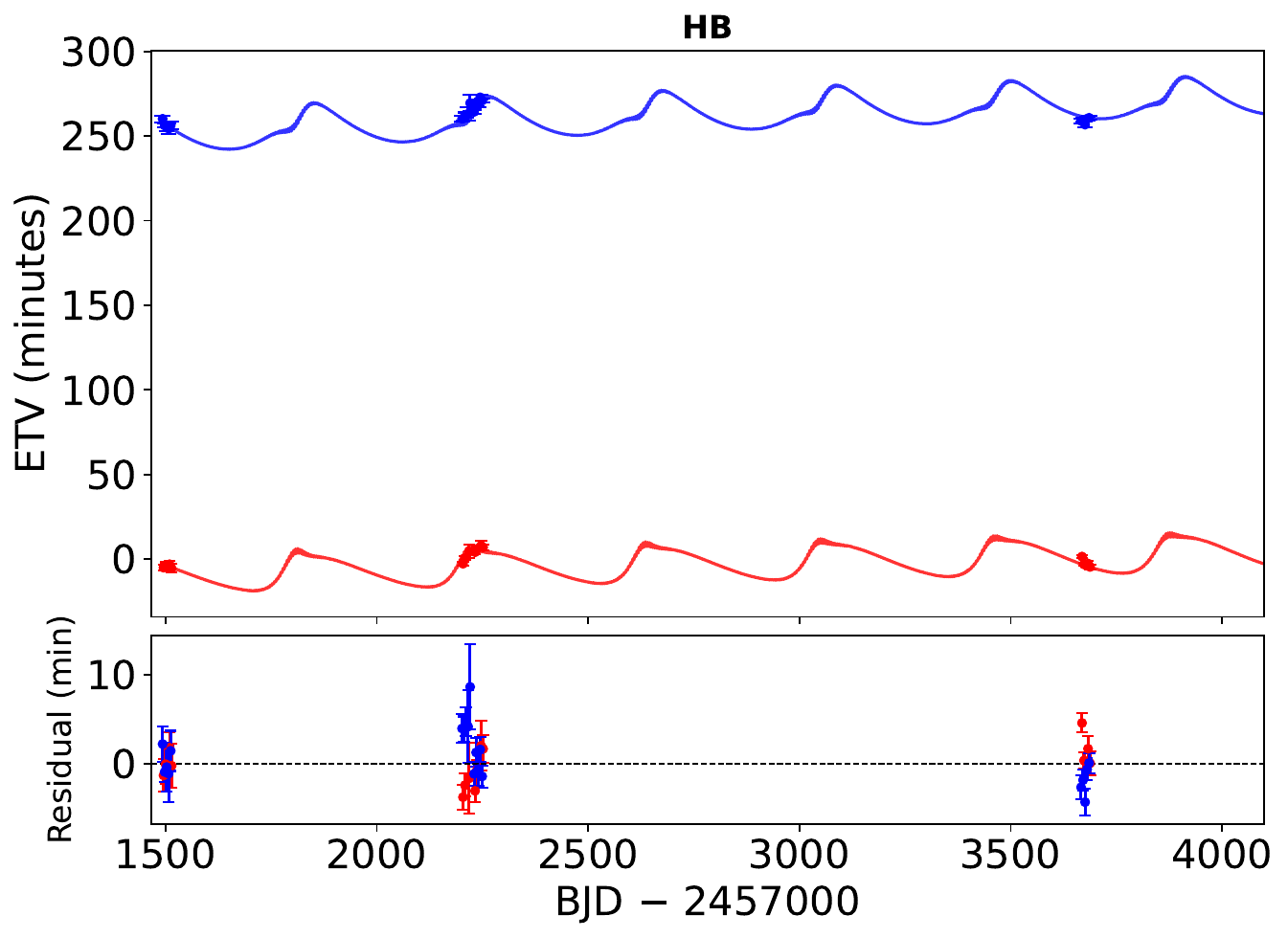}
    \caption{ ETV curves for RGB-s (top panel) and HB-s (bottom panel) from BJD 2458466.00 to 2460747.00 with primary measured ETVs (red circles), secondary measured ETVs (blue circles), primary model curve (red line), and secondary model curve (blue line).}
\label{fig:etv}
\end{figure}

We show the ETVs (points) and model fit (lines) of the EB primary (red) and secondary (blue) in Figure \ref{fig:etv}, with the RGB-s in the top panel and HB-s in the bottom panel.  The ETV data points are given by the calculated eclipse time subtracted from the observed eclipse time (O-C), extracted by the method discussed in Section \ref{sec::etv_model} and provided in Table \ref{tab:etv_data}.  The curve is generated by the analytical expressions given by \citet{2015MNRAS.448..946B}, with LTTE, dynamical, apsidal, and polynomial effects included.  The slightly differing shapes of the ETV curves for HB-s and RGB-s (particularly evident in the secondaries of each) are driven by the dynamical effects of the mass differences.

\begin{figure}
   \centering
    \includegraphics[width=1.\columnwidth]{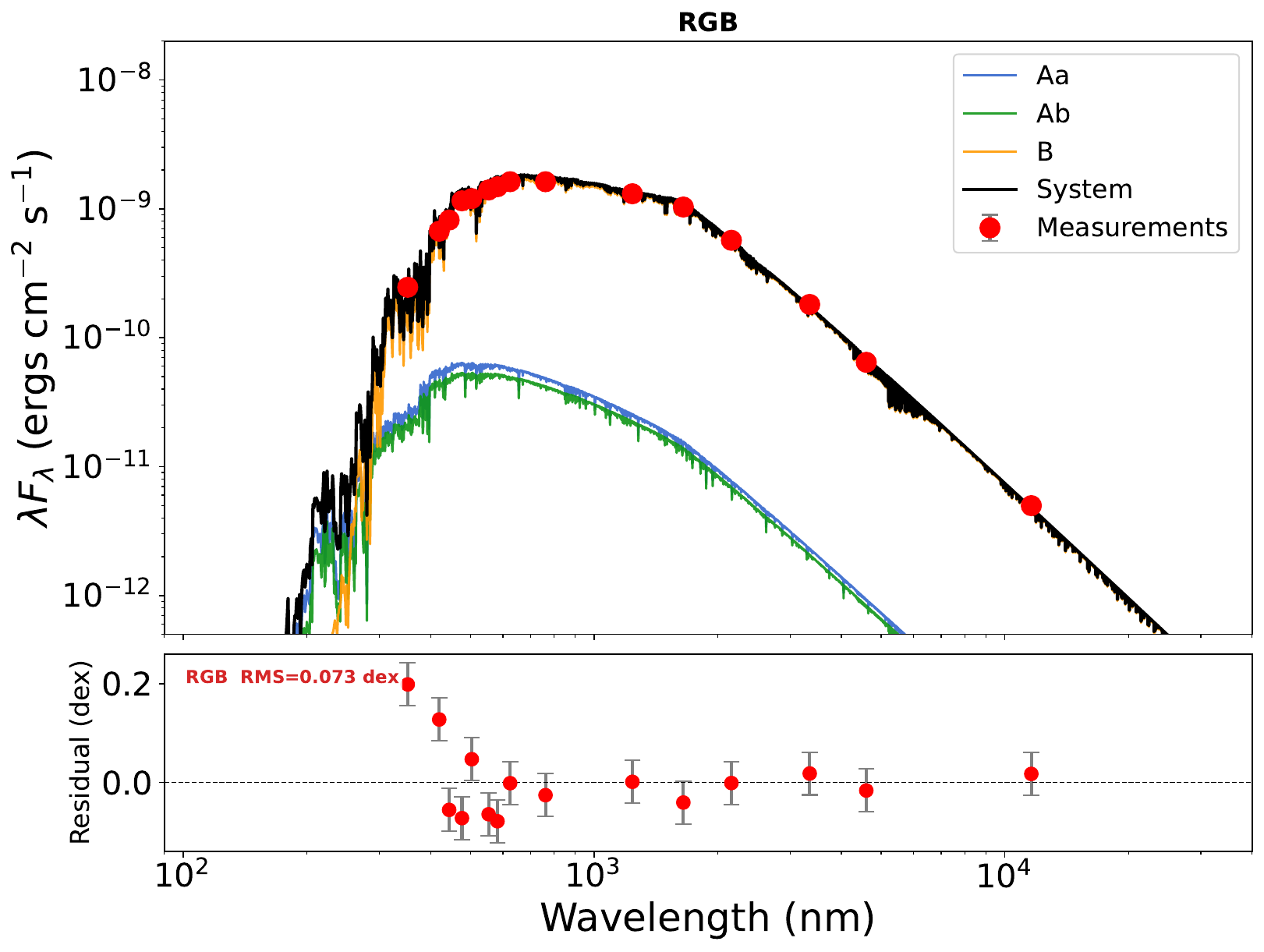}
    \includegraphics[width=1.\columnwidth]{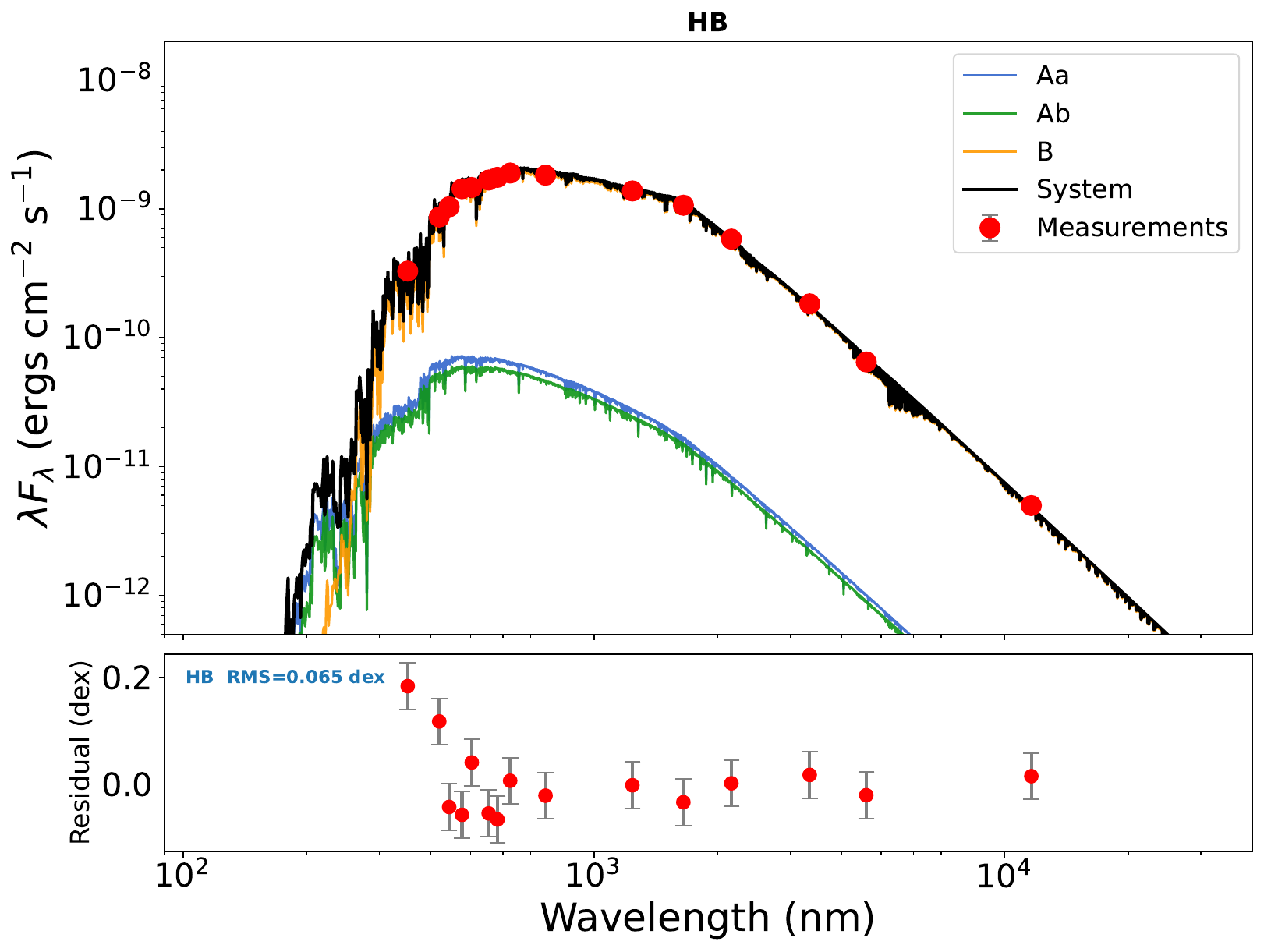}
    \caption{TIC 295741342 model SED fit at the star for RGB-s (top panel) and HB-s (bottom panel).  The model curves in both panels are nearly identical.  The binary components Aa and Ab are shown in blue and green, respectively.  The tertiary is shown in orange, with the total SED in black.  The total and the tertiary overlap in both solutions as the SED is dominated by the giant.}
    \label{fig:sed}
\end{figure}

The SED fits at the star are shown in Figure \ref{fig:sed}, with RGB-s in the top panel and HB-s in the bottom panel.  As can be expected, the giant completely dominates the SED in both cases, with only a small contribution from the stars in the binary.  To gather the data needed for this fit, we used the VizieR SED photometry viewer \citep{vizier2000}, finding the fluxes given in Table \ref{tab:sed_data}.  The SED model is created from the BT-Settl model atmospheres \citep{2014IAUS..299..271A} as discussed in Section \ref{sec::sed_model}.  Extinction is applied via the formulations of \citet{1989ApJ...345..245C}.

\begin{deluxetable}{lccrr}
\tablewidth{0pt}
\tablecaption{Photometric measurements used for SED fitting.\label{tab:sed_data}}
\tablehead{
\colhead{Band} & \colhead{$\lambda_{\mathrm{eff}}$} & \colhead{$\lambda_{\mathrm{range}}$} & \colhead{$F_\lambda$} \\[-6pt]
\colhead{} & \colhead{(nm)} & \colhead{(nm)} & \colhead{(erg\,s$^{-1}$\,cm$^{-2}$\,\AA$^{-1}$)}
}
\startdata
SDSS $u$          & 352   & 330--390     & $5.665 \times 10^{-14}$ \\
Hipparcos $B_T$   & 420   & 380--460     & $1.327 \times 10^{-13}$ \\
Johnson $B$       & 444   & 380--500     & $1.550 \times 10^{-13}$ \\
Pan-STARRS $g$    & 477   & 420--540     & $2.067 \times 10^{-13}$ \\
Gaia $G_{\mathrm{BP}}$    & 504   & 400--670     & $2.057 \times 10^{-13}$ \\
Johnson $V$       & 554   & 490--610     & $2.220 \times 10^{-13}$ \\
Gaia $G$          & 582   & 400--900     & $2.255 \times 10^{-13}$ \\
SDSS $r$          & 625   & 570--680     & $2.320 \times 10^{-13}$ \\
Gaia $G_{\mathrm{RP}}$    & 762   & 630--950     & $1.957 \times 10^{-13}$ \\
2MASS $J$         & 1240  & 1125--1375   & $1.017 \times 10^{-13}$ \\
2MASS $H$         & 1650  & 1525--1780   & $6.115 \times 10^{-14}$ \\
2MASS $K_s$       & 2160  & 2030--2320   & $2.600 \times 10^{-14}$ \\
WISE $W1$         & 3350  & 3000--3800   & $5.369 \times 10^{-15}$ \\
WISE $W2$         & 4600  & 4100--5200   & $1.394 \times 10^{-15}$ \\
WISE $W3$         & 11600 & 8000--16000  & $4.285 \times 10^{-17}$
\enddata
\tablecomments{Fluxes are observed values prior to extinction correction. The wavelength range indicates the passband width used for computing the mean model flux in each band. A 10\% uncertainty is adopted for all bands.}
\end{deluxetable}

\section{Discussion}
\label{sec::discussion}

The model fits presented in Section \ref{sec::results} agree nicely with the measured data in all aspects for both RGB-s and HB-s.  In constraining the system parameters, the thorough spectroscopy provided by the dedicated efforts of the TRES team formed an excellent complement to the {\em TESS} light curve showing the EB and the outer eclipse.  The shape of the outer eclipse self-confirms the system as a triple with its unique structure.  As we noted in Section \ref{sec::tess} and showed in Figure \ref{fig::zoom}, the outer eclipse by itself provides clues to the system structure, indicating that (i) the tertiary star is a giant, (ii) one star of the EB provides 2.3\% of the system light in the {\em TESS} band, and (iii) the other star of the EB provides 2.7\% of the system light in the {\em TESS} band.  The interesting shape requires a thorough explanation which we provide here in addition to an instructive figure.

\subsection{The Outer Eclipse}

\begin{figure*}
    \centering
    \includegraphics[width=.94\textwidth]{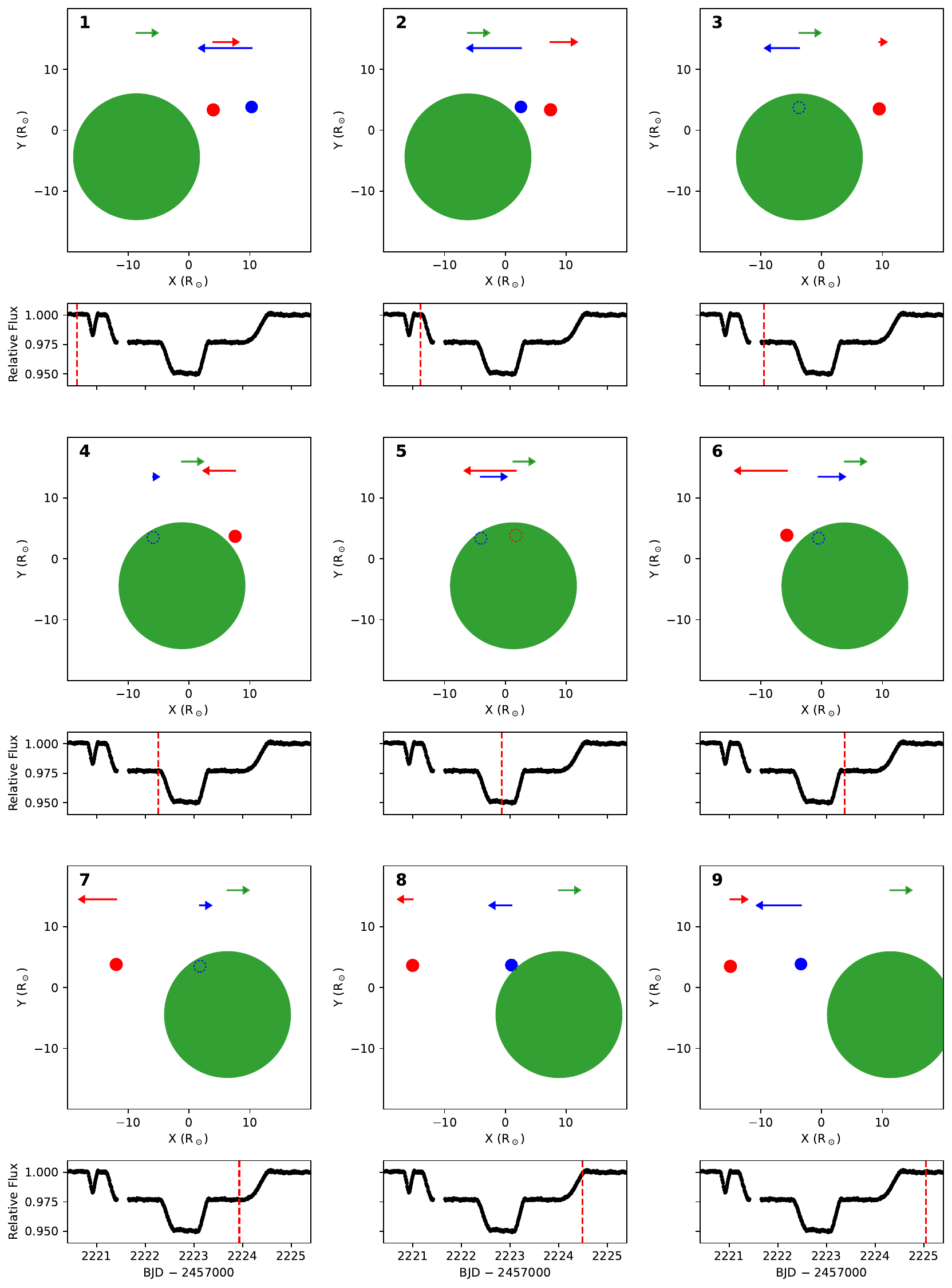}
    \caption{A detailed view of the HB solution outer eclipse event over nine timesteps.  ({\em upper panels}) View from the negative $Z$ direction of stars Aa (red), Ab (blue), and B (green). When stars Aa and Ab pass behind B, we show their positions as dashed outlines for clarity. ({\em lower panels}) {\em TESS} lightcurve data (black dots) for the duration of the outer eclipse event.  The dashed vertical red line shows the time corresponding to the positions of the stars in the upper panels.}
\label{fig:outer}
\end{figure*}

In Figure \ref{fig:outer}, we show a step-by-step explanation of the mechanics of the outer eclipse at nine different stages.  Given that our purpose here is to demonstrate the process by which the shape of the outer eclipse is formed, we show this only for HB-s to avoid a redundant figure since RGB-s is a nearly identical process. Star Aa is shown in red, star Ab in blue, and star B in green.  The position in the light curve at each stage is shown by the dashed red vertical line in the lower panel.  Arrows corresponding to the color of each star are at the top of the plot, with the length of the arrow representing its normalized speed in the $x$ direction.  Moving from top left to bottom right, we describe each stage:
\begin{enumerate}
    \item The EB (stars Aa and Ab) is moving to the left and the tertiary (star B) is moving to the right.  There are no eclipses at this time.
    \item The EB has undergone an eclipse since (1), with Ab nearly intersecting B.
    \item Ab is completely occulted by B, forming the initial ``shoulder'' of the outer eclipse, while the primary remains unobscured.  Aa and Ab are moving in opposite directions.
    \item Ab is still completely occulted by B, but Aa is moving quickly to the left, now nearly intersecting B.
    \item Both Aa and Ab are now occulted by B, forming the ``head'' of the outer eclipse.  Aa and Ab are moving towards each other and are nearing an eclipse, but this will not be visible since both are behind B.
    \item Aa and Ab have undergone an unobserved eclipse and are now moving in opposite directions.  Aa has emerged from behind B, forming the latter ``shoulder'' of the outer eclipse with Ab still obscured.
    \item Aa and Ab are still moving in opposite directions, with Aa moving to the left and Ab moving to the right.  B is also moving to the right, so the relative motions cause Ab to remain fully obscured by B.
    \item Aa and Ab are now both moving to the left.  Ab has almost completely emerged from behind B while B moves to the right.
    \item Aa and Ab are both unobscured.  They are moving towards each other and will soon undergo an eclipse.  B has moved to the right and will not obscure either star again until a full revolution of the outer orbit.
\end{enumerate}

Naturally, the question arises of whether this specific eclipse pattern will repeat in the future.  In order to determine the future behavior, we simulated our models forward for several years and examined the outer eclipse events.

\subsection{Past \& Future Outer Eclipses}

\begin{figure*}
   \centering
    \includegraphics[width=0.32\linewidth]{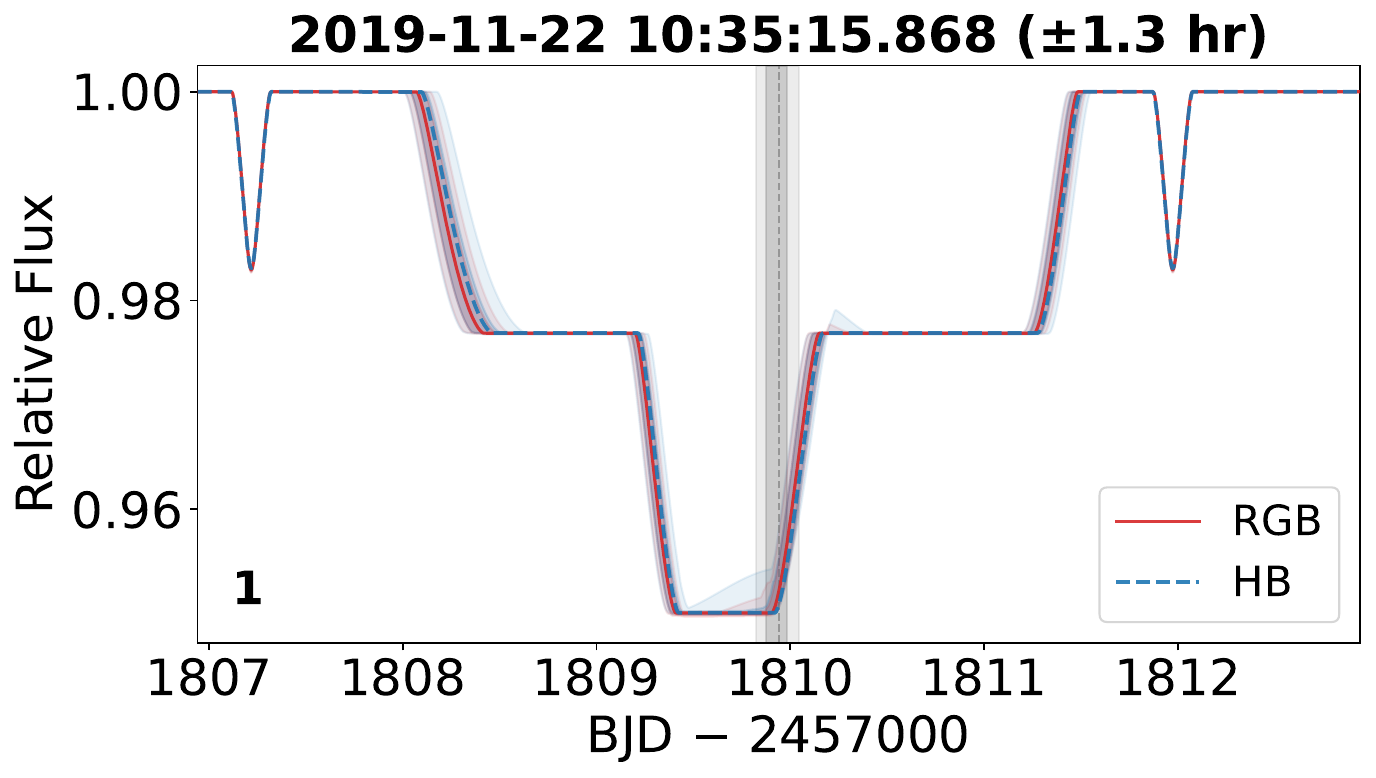}
    \includegraphics[width=0.32\linewidth]{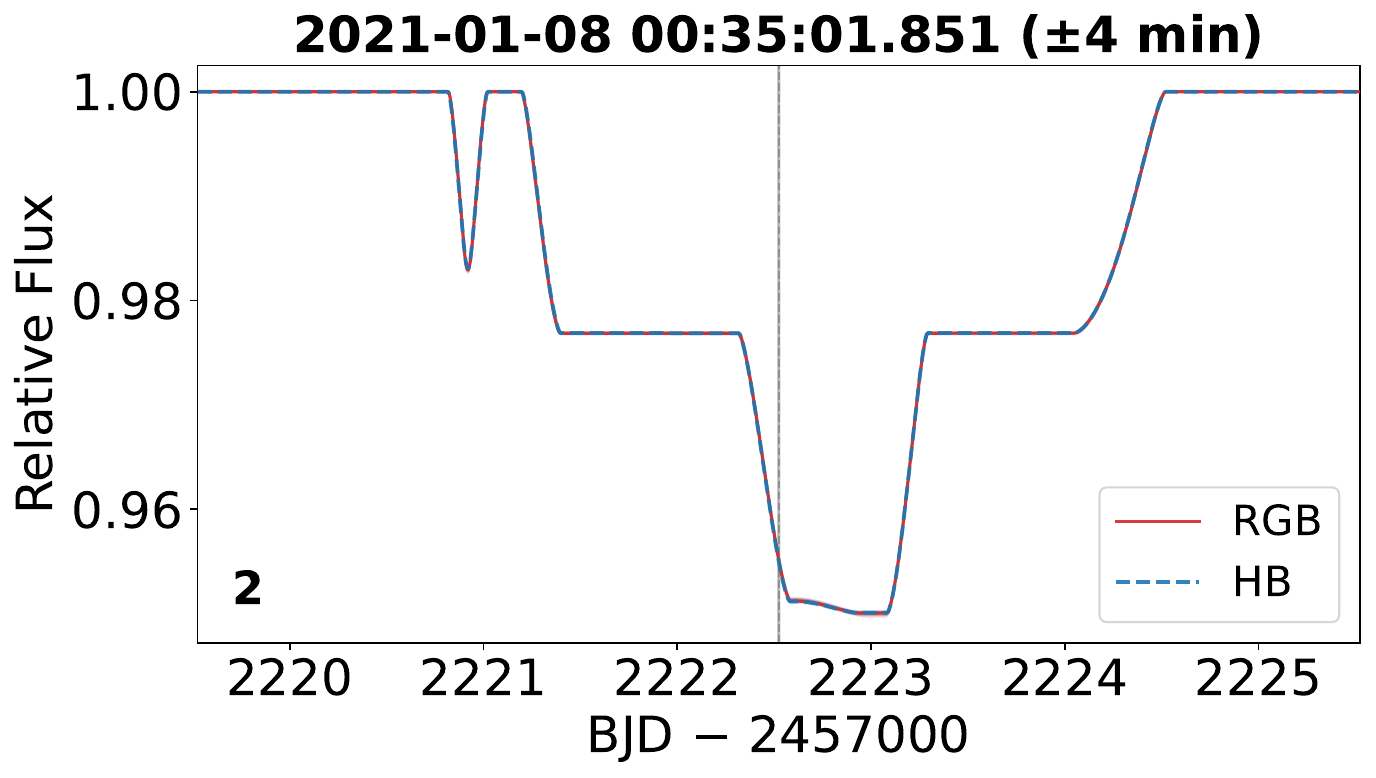}
    \includegraphics[width=0.32\linewidth]{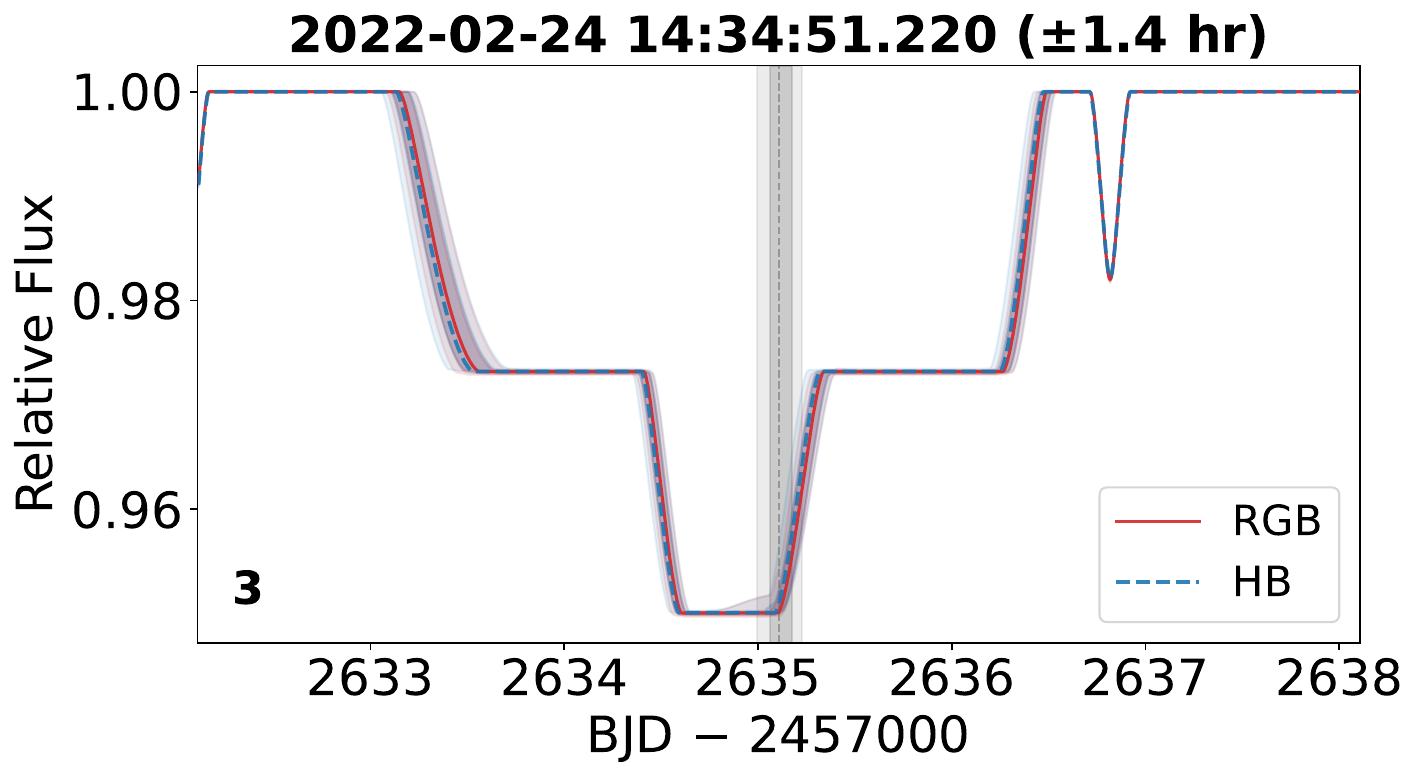}
    \includegraphics[width=0.32\linewidth]{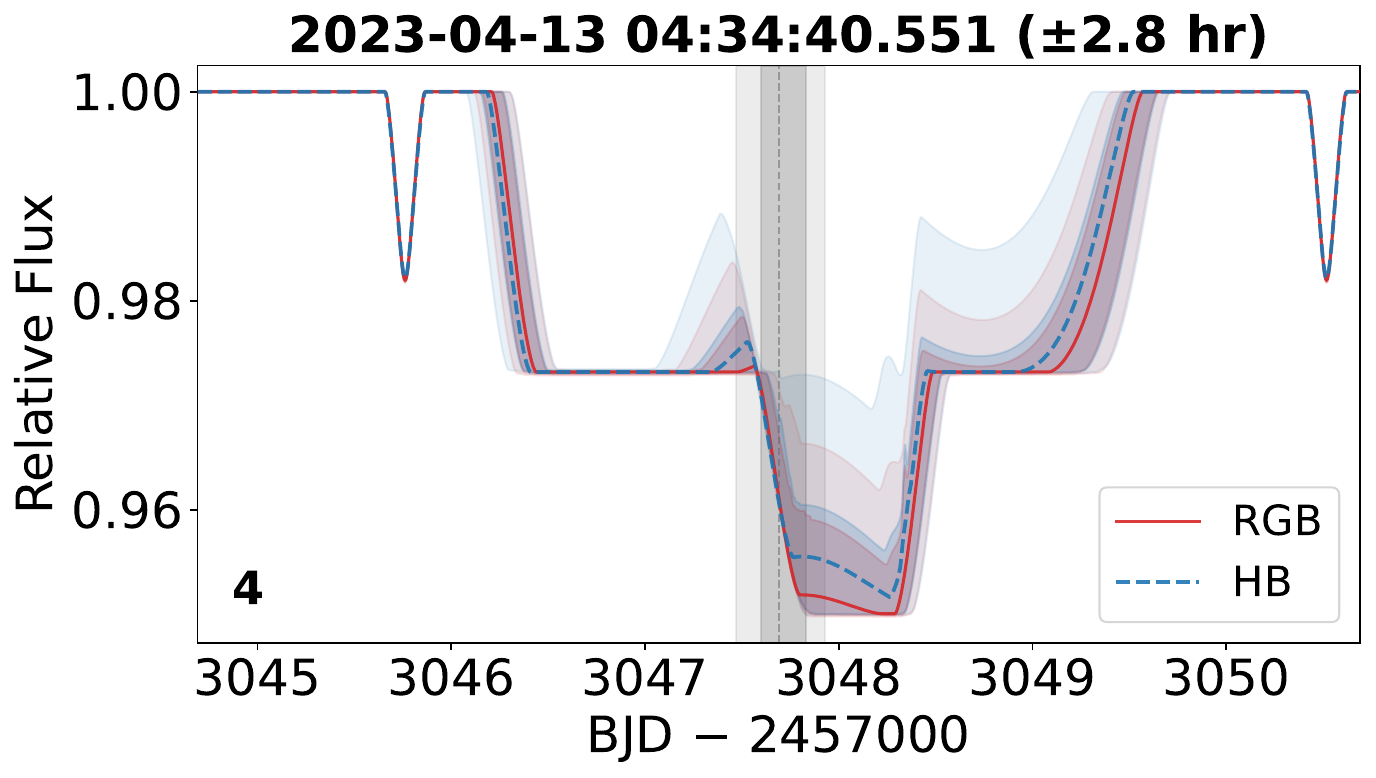}
    \includegraphics[width=0.32\linewidth]{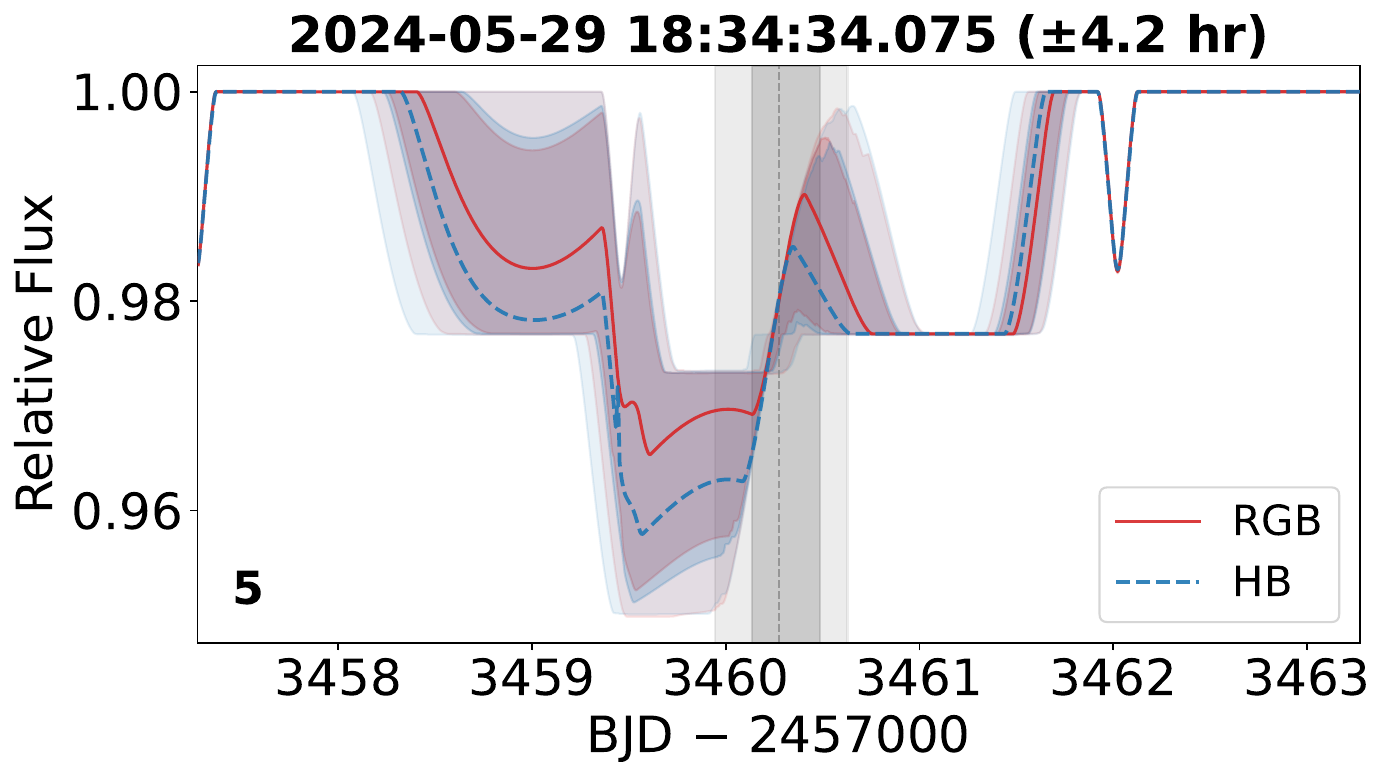}
    \includegraphics[width=0.32\linewidth]{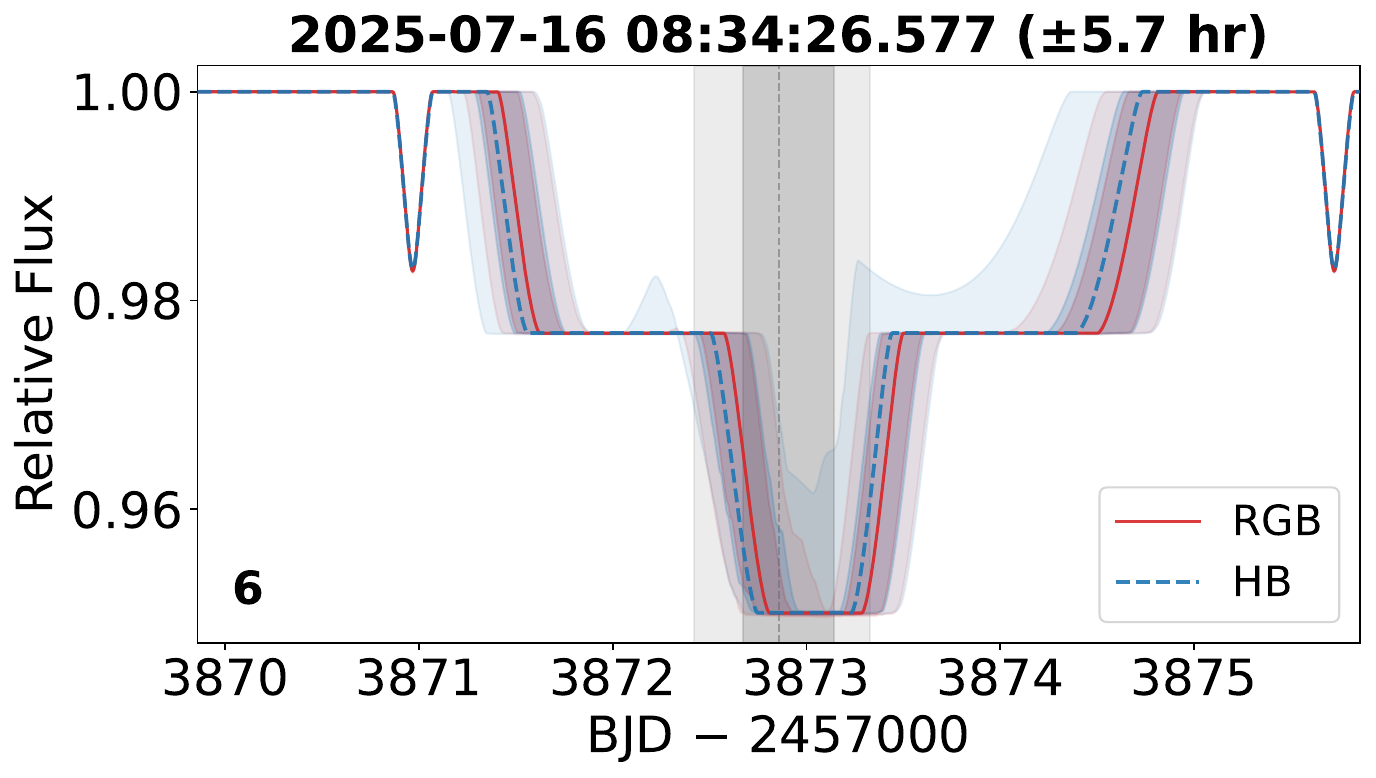}
    \includegraphics[width=0.32\linewidth]{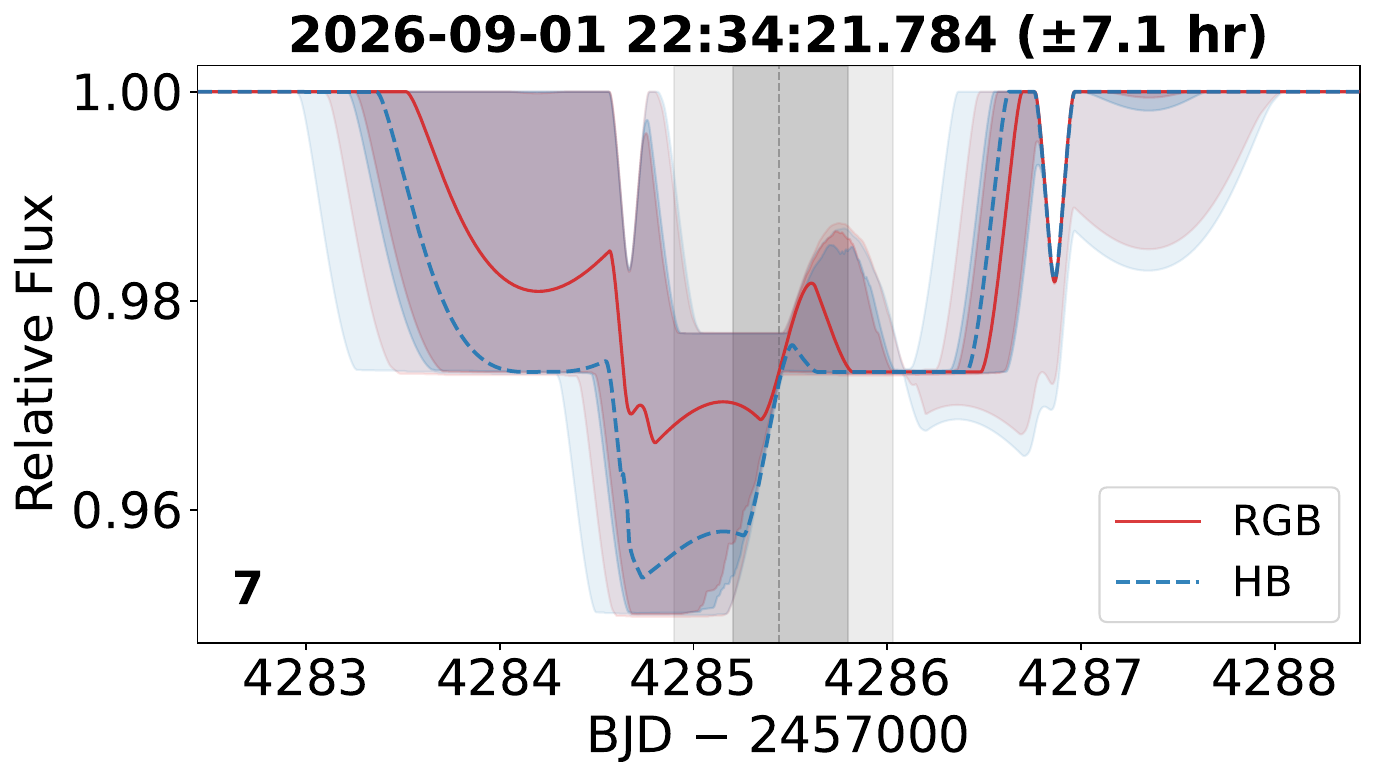}
    \includegraphics[width=0.32\linewidth]{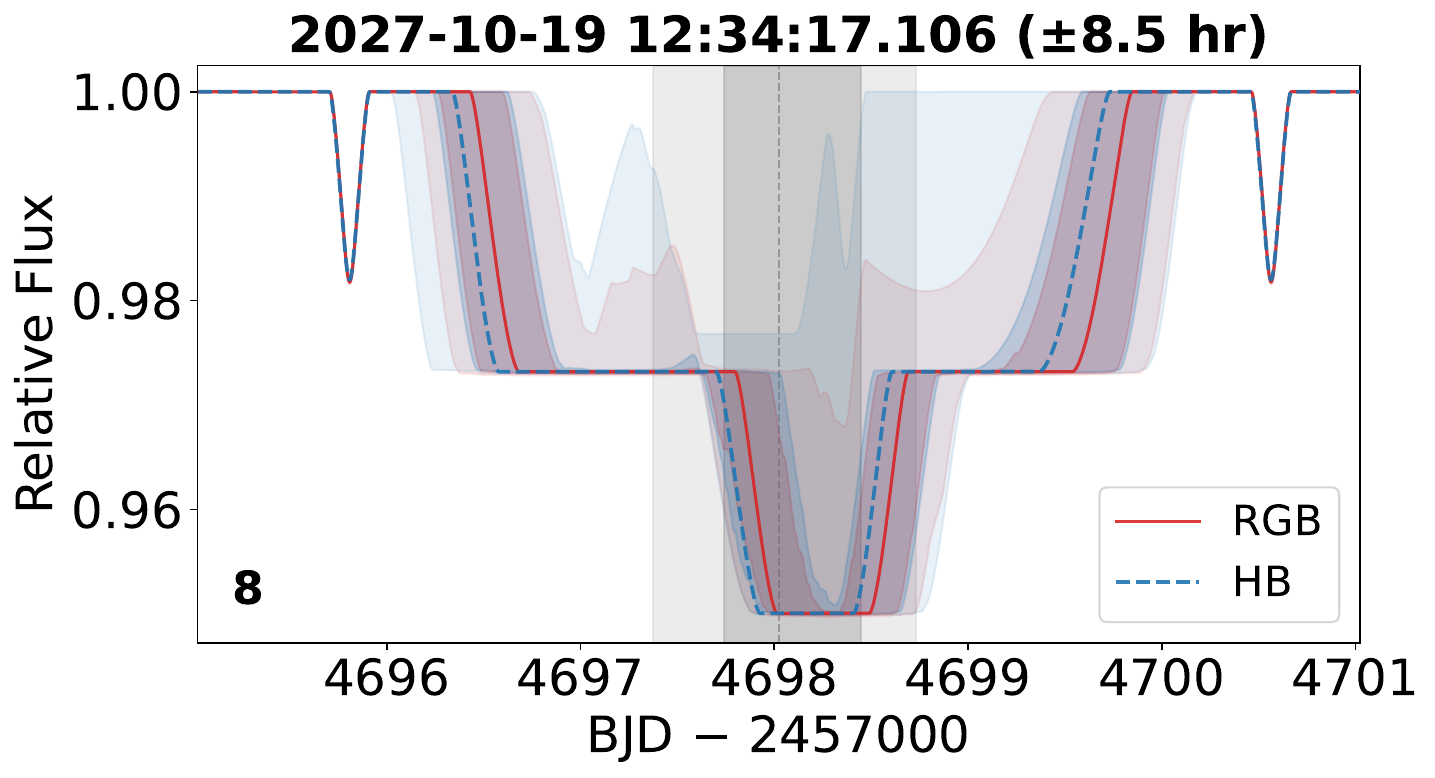}
    \includegraphics[width=0.32\linewidth]{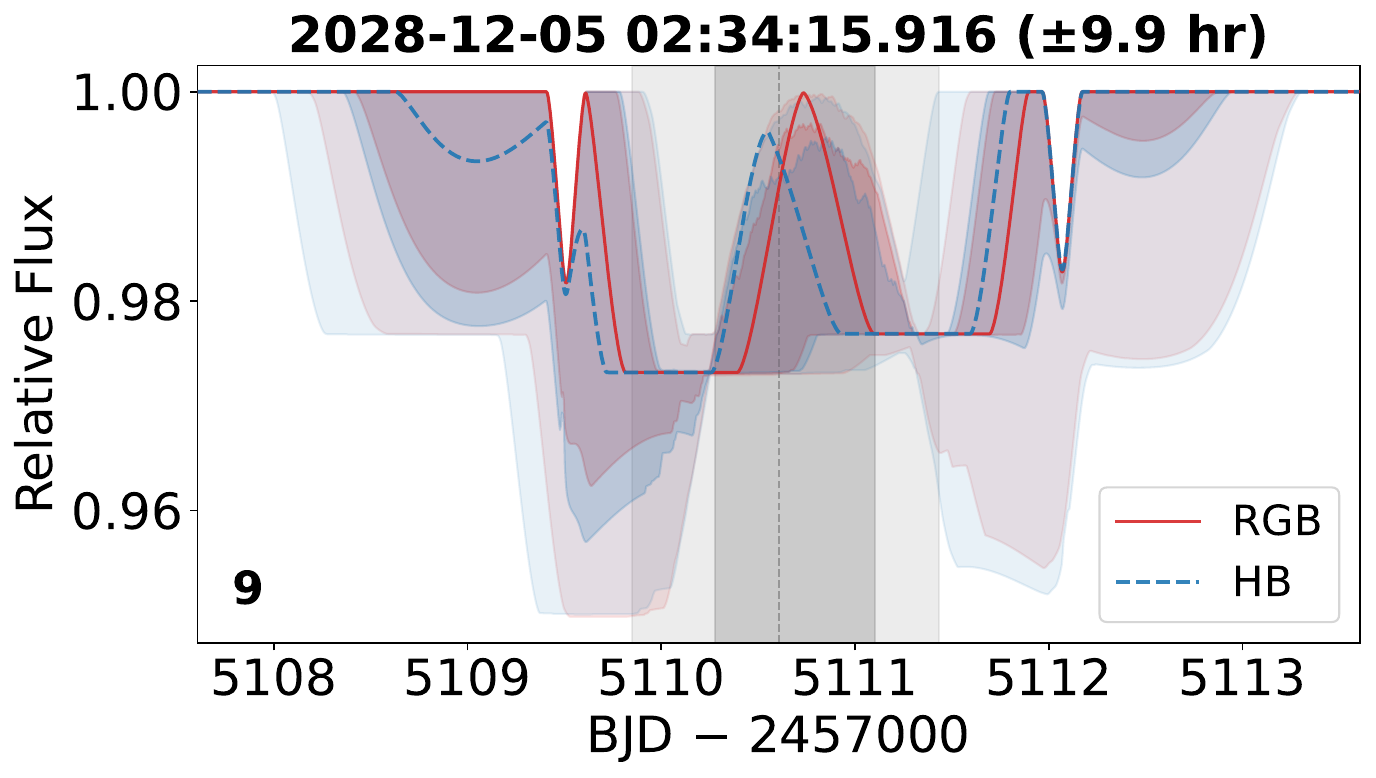}
    \caption{Lightcurves of outer eclipses for the duration of the model fit and future projections.  The best model flux is shown as a red line, with 1$\sigma$ and 2$\sigma$ uncertainty bands created from 1,000 random parameter sets in the posterior.  The vertical gray line is the time of the closest approach between the tertiary and the center of mass of the binary in the XY-plane with 1$\sigma$ and 2$\sigma$ bands.  The title of each panel gives the midpoint UTC date/time with uncertainties.  Eclipse 1 was the first to occur during the simulation but was unobserved.  Eclipse 2 is the modeled eclipse from Sector 33, hence the visually imperceptible uncertainty. Eclipses 3-6 already occurred but were not observed. As of the submission date of this paper, eclipses 7-9 have not yet occurred.}
    \label{fig:future_eclipses}
\end{figure*}

The lightcurves of the outer eclipses for the duration of the models and into the future are shown in Figure \ref{fig:future_eclipses}.  The flux of the best RGB-s solution is shown as a solid red line, with 1$\sigma$ and 2$\sigma$ bands in shades of red created by randomly sampling 1,000 parameter sets from the posterior and generating the full light curve from each.  The flux of the best HB-s solution is shown as a dashed blue line with the 1$\sigma$ and 2$\sigma$ bands in shades of blue created in the same manner.  A gray vertical line indicates the time of the minimum distance between the tertiary and the center of mass of the binary in the XY-plane (also provided with 1$\sigma$ and 2$\sigma$ bands), which is also the time given in the title of each plot.  A reference number for each outer eclipse is given in the lower left of each panel.  Eclipse 1 was the first to have occurred during the course of our simulation, but was not observed.  Eclipse 2 was observed in {\em TESS} sector 33 and modeled, hence the small uncertainty.  Eclipses 3-6 have already occurred, but have not been observed.  The next eclipse to occur as of publication of this paper is eclipse 7, with a midpoint on September 1, 2026.  The shape of a future eclipse, as well as its timing, would severely constrain the system parameters.  Therefore, we suggest that an observation of the next outer eclipse (with a minimum $\pm$3 day window around the September 1, 2026, 22:34:22 UTC midpoint to ensure observation of the full event) incorporated into a new model fit would substantially reduce the model uncertainty such that future eclipses could be predicted accurately for several years.

\subsection{Formation \& Evolution}
\label{sec:formation_evolution}

The most likely formation mechanism for the TIC 295741342 system is disk fragmentation.  Indeed, \citet{2016Natur.538..483T} provided direct observational evidence of the formation of a triple star system from the fragmentation of a protostellar disk, initially at a wide orbit with inward orbital migration. The fragments would accrete mass as the disk dissipates and evolve into a stable orbit after an initial period of instability \citep{2016ARA&A..54..271K}. \citet{2017ApJ...844..103T} suggested that the orbital configurations of coplanar triples (i.e. with small mutual inclinations) with outer orbits $<50$ au are likely shaped by gas dissipation within the disk.  The TIC 295741342 configuration in either solution falls well within this regime.

\begin{figure*}
   \centering
    \includegraphics[width=0.99\linewidth]{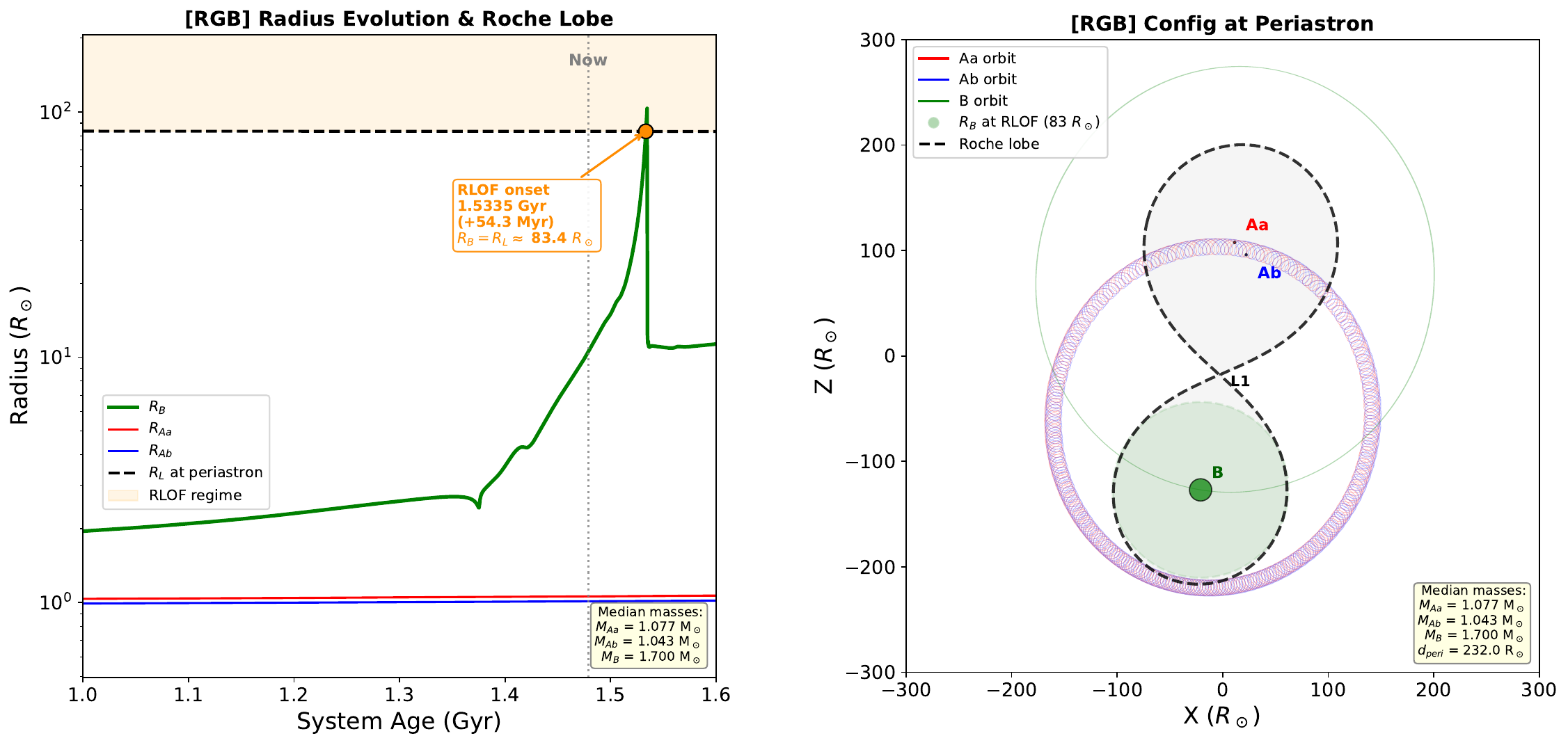}
    \includegraphics[width=0.99\linewidth]{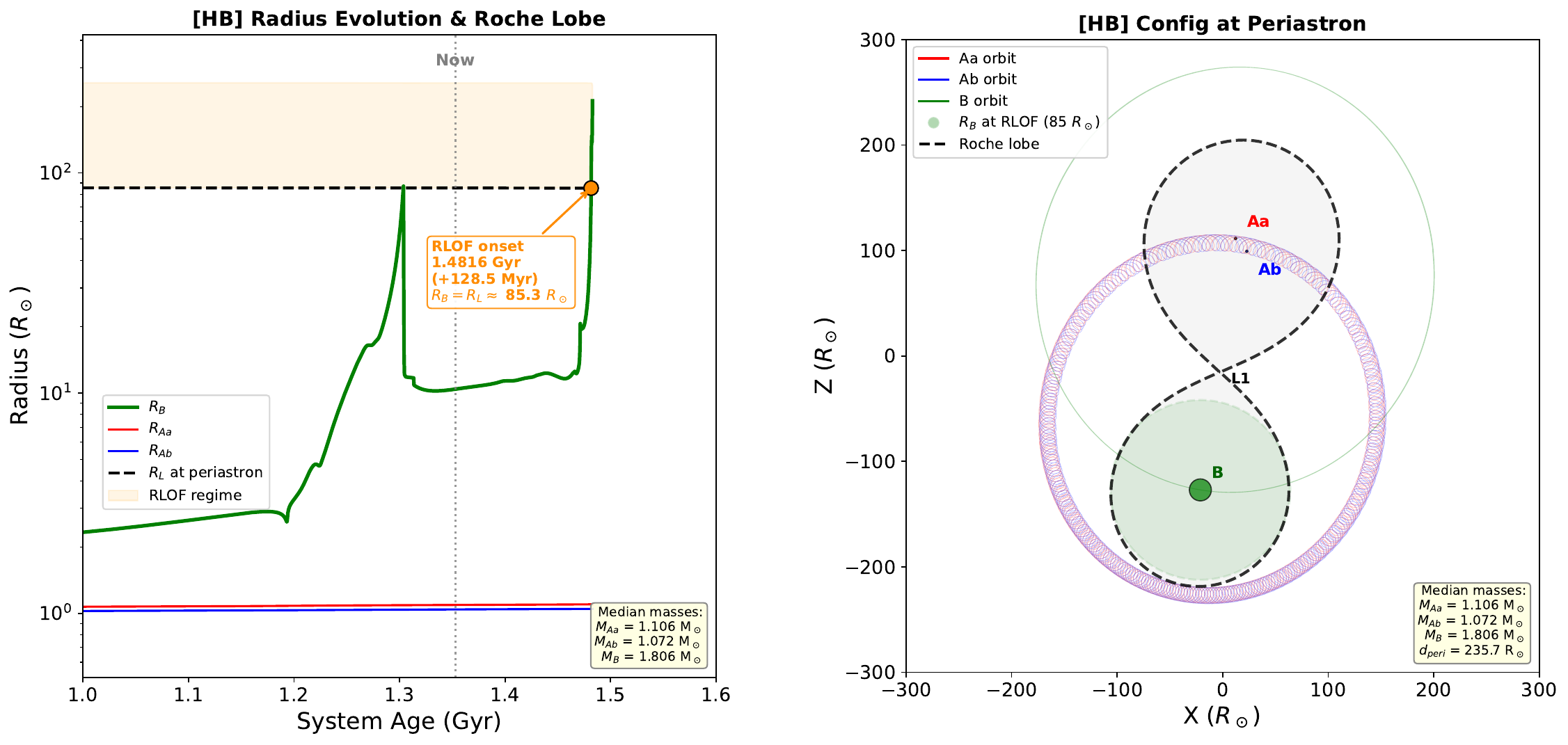}
    \caption{({\em left panels}) Visualizations of the radii of the stars in TIC 295741342 as the system evolves. For RGB-s in the top row, the radii of stars Aa and Ab will slightly decrease over the duration in both cases.  The radius of star B (green) will grow rapidly as it ascends the RGB.  In \rgbrlof Myr, star B will overflow its Roche lobe with a radius of \rgbrlofrad R$_\odot$.  For HB-s in the bottom row, the system has survived star B's full ascent of the RGB and transition to the HB.  The radius of star B will grow slowly for the next $\sim$100 Myr until it begins to ascend the AGB. In 129 Myr, the radius of star B will grow to \hbrlofrad R$_\odot$ and begin RLOF.  ({\em right panels}) View of the system from the positive Y direction for RGB-s (top right) and HB-s (bottom right) showing the orbital configuration at periastron (red, blue, and green lines and circles for the orbits and positions of stars Aa, Ab, and B, respectively).  The dashed black line shows the Roche lobe, with the shaded green area showing star B at RLOF.}
    \label{fig:evolution}
\end{figure*}

As for evolution in both RGB-s and HB-s, the disparity in the radius of the tertiary vs the stars in the binary is the most interesting quality of the system.  To examine the nature of this relationship in the radii, we used the MIST isochrones (as discussed in Section \ref{sec::mist_model}) to trace the system from its origins and predict its future.  With the initial masses and [Fe/H] known from our model fit, we can follow the evolution of each star through interpolation of the isochrones.  We provide a visualization of the evolution of the radii in the left panels of Figure \ref{fig:evolution}, with RGB-s in the top left panel and HB-s in the bottom left panel.  These are plotted on identical x-axes for visual comparison purposes with system age from 1.0 - 1.6 Gyr.  The right panels show a view of the system from the positive Y direction showing the orbital configuration at periastron (red, blue, and green lines for Aa, Ab, and B, respectively).  The dashed black line shows the Roche lobe, with the shaded green area showing B at Roche lobe overflow (RLOF).  RGB-s is shown in the top right panel and HB-s is shown in the bottom right panel.

In the left panels, note the dashed, gray vertical line indicating the current age of the system.  In RGB-s (top left panel), star B is ascending the RGB and will see rapid growth in its radius over the next \rgbrlof Myr until it reaches RLOF at 83 R$_\odot$.  In HB-s (bottom left panel), the system survived the evolution of star B through the RGB and the transition to the HB (discussed further in the following paragraph).  The radius of star B will slowly increase over the next $\sim$100 Myr until it rapidly grows as star B ascends the Asymptotic Giant Branch (AGB).  Star B will overflow its Roche lobe in 129 Myr when its radius reaches 85 R$_\odot$.

In HB-s, we note that the MIST models suggest that B barely reaches RLOF at the tip of the RGB.  While this can result in mass transfer and other physical changes to the system orbit \citep{Toonen2016}, we of course cannot know what happened at this stage of evolution and, in the HB solution, we only surmise that the system survived in its current form.  We assess two possible scenarios at the tip of the RGB that allow for HB-s: (i) Given the uncertainties inherent in the MIST models as well as uncertainties in the fitted masses and radii, star B failed to actually fill its Roche lobe at the tip of the RGB, or (ii) if star B did barely fill its Roche lobe (as the bottom left panel of Figure \ref{fig:evolution} suggests), it is possible the mass transfer was stable and that very little mass was transferred before the star B descended from the tip of the RGB.  If, on the other hand, star B did barely fill its Roche lobe and the subsequent mass transfer was not stable\footnote{In this regard, we note that $q_{\rm out} = 0.83 \pm 0.01$.} (depending on how much mass was ejected and with what specific angular momentum) and a common envelope ensued, no HB-s would be possible.

In a series of simulations, \citet{2022ApJS..259...25H} found that, of triple star systems that undergo RLOF, $\sim$64\% proceed to stable mass transfer and $\sim$36\% proceed to common envelope evolution with one or more mergers.  A variety of possible outcomes involving mergers and/or ejections results from the latter, and we refer the reader to \citet{2021MNRAS.500.1921G} for a thorough discussion of all possibilities.

\subsection{Evaluation of the Two Solutions}
\label{sec::sol_eval}

Both RGB-s and HB-s produce equally good fits to the observational measurements (RVs, lightcurve, SED, eclipse times), making the selection of one over the other as the physically realized solution particularly difficult.  As such, we evaluated the plausibility of the solutions as a function of dwell time, i.e., how long can the system exist with reasonable parameters to produce the provided fits in either solution.  We randomly selected $10^5$ parameter sets from the posterior for both RGB-s and HB-s and ran each through the following process:
\begin{enumerate}
    \item Define the tolerance window for a viable physical system as $R_B\pm\sigma(R_B)$ from Table \ref{tab:rgb_mcmc_results} or Table \ref{tab:hb_mcmc_results} for RGB-s or HB-s, respectively.
    \item Select the initial mass of star B, log(age), and [Fe/H] from the posterior sample and define the starting point of the simulation as $t_0=\text{log(age)}$.  
    \item Evolve star B forward in time from $t_0$ in increments of $\delta_{\text{log(age)}}=1\times10^{-4}$ until the radius is $\geq R_B+\sigma(R_B)$.  The time of the final step is $t_\text{forward}$.
    \item Go back to $t_0$ and evolve star B backward in time in increments of $\delta_{\text{log(age)}}=1\times10^{-4}$ until the radius is $\leq R_B-\sigma(R_B)$.  The time of the final step is $t_\text{backward}$.
    \item Calculate the evolutionary dwell time in the tolerance window as $10^{t_\text{forward}}-10^{t_\text{backward}}$.
\end{enumerate}

\begin{figure*}
   \centering
    \includegraphics[width=0.99\linewidth]{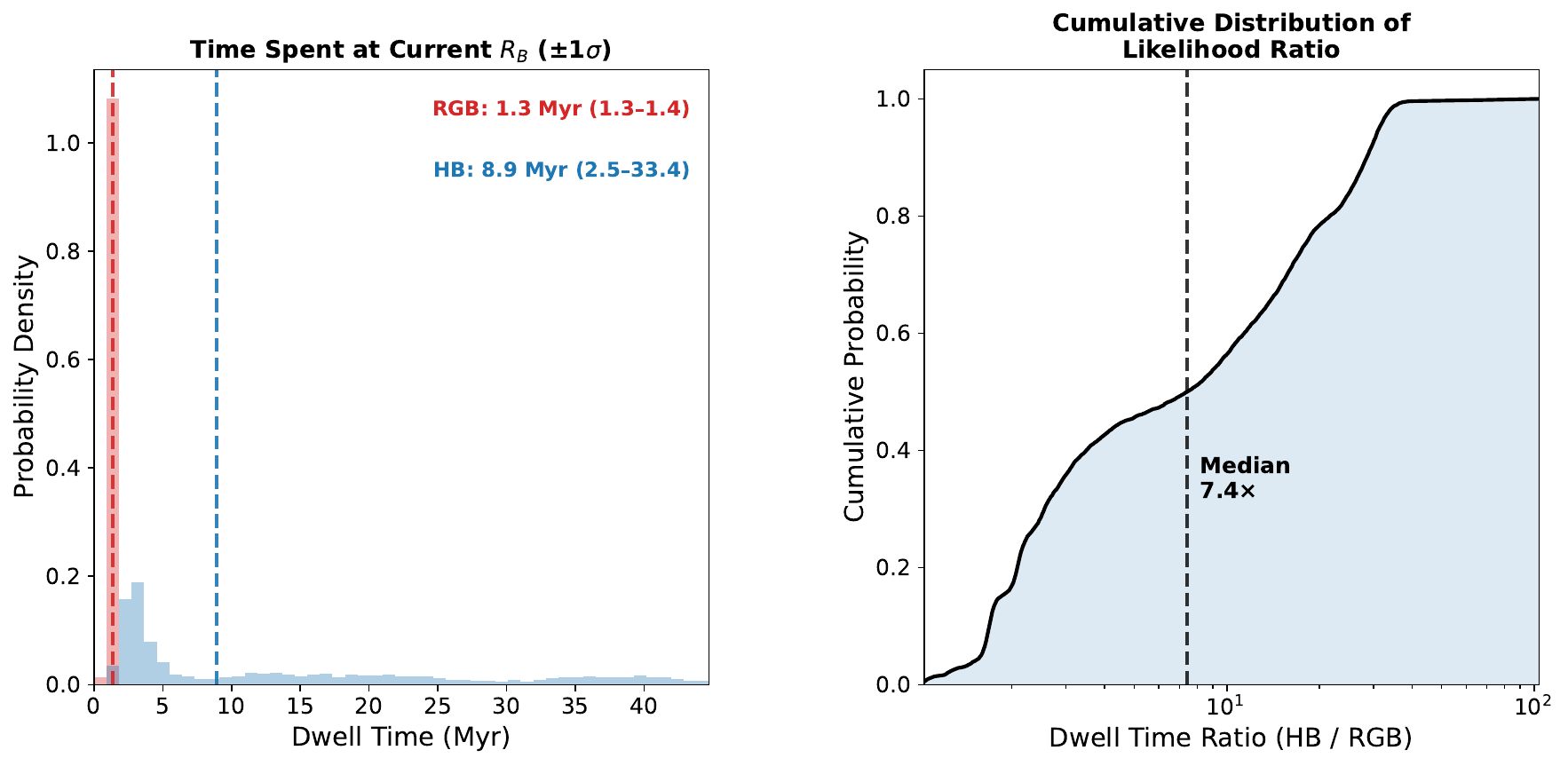}
    \caption{({\em left panel}) Histogram of $10^5$ dwell times calculated in the process described in Section \ref{sec::sol_eval} with parameter sets randomly sampled from the posteriors RGB-s (red) and HB-s (blue). Median dwell times and the 1$\sigma$ range are noted for both solutions in the upper right, with median values shown as dashed vertical lines.  ({\em right panel}) CDF of the HB/RGB dwell time ratio calculated from the same samples.  The median dwell time ratio is shown as the dashed vertical line.}
    \label{fig:dwell}
\end{figure*}

We accumulated and compared the dwell times, resulting in the distributions shown in Figure \ref{fig:dwell}.  In the left panel, we show a histogram of the dwell times, with HB-s in blue and the RGB-s in red.  It can be easily seen that the dwell time RGB-s is rather small, with a 1$\sigma$ range of 1.3-1.4 Myr, whereas the 1$\sigma$ range of dwell times for HB-s is 2.5-33.4 Myr.  This is, of course, due to the fact that the radius of star B is rapidly increasing on the ascent of the RGB, whereas on the HB the rate of change is much slower.

In the right panel, we show the CDF of the dwell time ratio (HB/RGB).  The median of the distribution is shown at a $\sim7.4\times$ ratio, i.e., the time spent in a valid parameter space for HB-s is, at the median, $\sim7.4\times$ greater than in RGB-s.  Purely based on this metric, we are much more likely to observe the HB-s system physically realized than we are to observe RGB-s.  This result does not reflect on the merit of the RGB-s, which is perfectly reasonable.  Rather, in a purely probabilistic sense, it is substantially more likely that we are observing HB-s than RGB-s based on how long valid solutions for either can exist.  Future observations of the system may help to constrain physical parameters such that one solution can be definitively ruled out.  The one caveat to this is that we have no quantitative way to evaluate the probability that the system would survive the RGB phase without filling its Roche lobe and transferring a significant amount of mass.  In the meantime, however, we adopt the HB solution as our preferred model while acknowledging that the RGB solution remains a viable alternative.

\subsection{Similar Systems}

TIC 295741342 is one of a small population of known triply-eclipsing triple star systems with a giant tertiary.  In Table \ref{tab:giants} we compiled properties of the giant tertiaries of similar such systems discovered in {\em TESS}.  It is notable that TIC 295741342 has, by far, the lowest mutual inclination of any of the given systems with giant tertiaries.  In terms of other known triply-eclipsing triples, it joins several other near-coplanar systems such as EPIC 249432662 \citep{2019MNRAS.483.1934B} ($0.173\pm0.397$ deg), HD 144548 \citep{2015A&A...584L...8A} ($0.2\pm0.5$ deg), TIC 209409435 \citep{2020MNRAS.496.4624B} ($0.24\pm0.08$ deg), and TIC 278825952 \citep{2020MNRAS.498.6034M}($0.5\pm0.3$ deg).

\begin{deluxetable*}{lcccccccccl}
\tablecaption{{\em TESS} triply eclipsing triple star systems with evolved/giant tertiary components from uniform photodynamical analyses.\label{tab:evolved_tertiaries}}
\renewcommand{\arraystretch}{1.2}
\tablehead{
\colhead{TIC} & \colhead{$R_B$} & \colhead{$M_B$} & \colhead{$T_B$} & \colhead{$\log g_B$} & \colhead{Age} & \colhead{$P_{\rm out}$} & \colhead{$P_{\rm in}$} & \colhead{$i_{\rm mut}$} & \colhead{$e_{\rm out}$} & \colhead{Ref.} \\[-6pt]
\colhead{} & \colhead{($R_\odot$)} & \colhead{($M_\odot$)} & \colhead{(K)} & \colhead{(dex)} & \colhead{(Gyr)} & \colhead{(d)} & \colhead{(d)} & \colhead{(deg)} & \colhead{} & \colhead{}
}
\startdata
37743815  & 4.01  & 1.61 & 5434 & 3.44 & 2.4 & 68.8  & 0.907 & 1.8  & 0.361 & \citet{2022MNRAS.513.4341R} \\
42565581  & 8.41  & 2.26 & 5218 & 2.94 & 1.1 & 123.5 & 1.824 & 5.5  & 0.161 & \citet{2022MNRAS.513.4341R} \\
54060695  & 8.35  & 2.10 & 5085 & 2.92 & 1.0 & 60.8  & 1.061 & 3.2  & 0.015 & \citet{2022MNRAS.513.4341R} \\
242132789 & 12.22 & 1.54 & 4734 & 2.45 & 2.5 & 42.0  & 5.129 & 2.0  & 0.006 & \citet{2022MNRAS.513.4341R} \\
456194776 & 4.94  & 1.94 & 5920 & 3.34 & 1.4 & 93.9  & 1.719 & 1.5  & 0.288 & \citet{2022MNRAS.513.4341R} \\
99013269  & 11.51 & 2.37 & 4798 & 2.69 & 0.9 & 604.2 & 6.535 & 20.2 & 0.463 & \citet{2023MNRAS.521..558R} \\
280883908 & 13.17 & 1.78 & 4704 & 2.45 & 1.8 & 184.6 & 5.242 & 2.3  & 0.260 & \citet{2023MNRAS.521..558R} \\
294803663 & 13.36 & 2.78 & 4902 & 2.63 & 0.5 & 153.4 & 2.246 & 3.1  & 0.030 & \citet{2023MNRAS.521..558R} \\
332521671 & 8.37  & 1.94 & 4946 & 2.88 & 1.4 & 48.5  & 1.248 & 2.8  & 0.004 & \citet{2023MNRAS.521..558R} \\
287756035 & 6.75  & 1.13 & 4833 & 2.84 & 5.7 & 367.9 & 2.082 & 1.4  & 0.235 & \citet{2024AA...686A..27R} \\
323486857 & 7.54  & 1.59 & 5068 & 2.88 & 2.4 & 41.4  & 0.885 & 2.3  & 0.007 & \citet{2024AA...686A..27R} \\
403916758 & 6.93  & 1.75 & 5372 & 3.00 & 1.9 & 71.1  & 1.134 & 4.1  & 0.031 & \citet{2025AA...703A.153B} \\
\hline
295741342 (RGB) & \rgbrb & \rgbmb & \rgbteffb & \rgbloggb & \rgbageb & \rgbpab & \rgbpa & \rgbmutinctab & \rgbeab & This work \\
295741342 (HB) & \hbrb & \hbmb & \hbteffb & \hbloggb & \hbageb & \hbpab & \hbpa & \hbmutinctab & \hbeab & This work \\
\hline
\enddata
\tablecomments{$R_B$, $M_B$, $T_B$, and $\log g_B$ are the radius, mass, effective temperature, and surface gravity of the evolved tertiary star. Age is the system age derived from isochrone fitting. $P_{\rm out}$ and $P_{\rm in}$ are the outer and inner orbital periods. $i_{\rm mut}$ is the mutual inclination between the inner and outer orbital planes. $e_{\rm out}$ is the eccentricity of the outer orbit.  Uncertainties are omitted for the purpose of general comparison.  We present both the RGB-s and HB-s for TIC 295741342 to accommodate the degeneracy in the evolutionary state of the tertiary. However, only one state is physically realized.}
\label{tab:giants}
\end{deluxetable*}

\section{Summary}
\label{sec::summary}

We have presented the discovery and characterization of TIC 295741342, a triply-eclipsing triple star system with a giant tertiary. There are two distinct solutions to the system, separated by the evolutionary state of the tertiary.  In RGB-s, the eclipsing binary consists of two similar stars with masses of \rgbmaa M$_{\odot}$ and \rgbmab M$_{\odot}$ and radii of \rgbraa R$_{\odot}$ and \rgbrab R$_{\odot}$ in a \rgbpa-day orbit. The tertiary is a giant with a mass of \rgbmb M$_{\odot}$ and a radius of \rgbrb R$_{\odot}$ in a \rgbpab-day orbit with the binary. The system is near-perfectly coplanar, with a mutual inclination of just \rgbmutinc degrees.  In HB-s, the eclipsing binary consists of two similar stars with masses of \hbmaa M$_{\odot}$ and \hbmab M$_{\odot}$ and radii of \hbraa R$_{\odot}$ and \hbrab R$_{\odot}$ in a \hbpa-day orbit. The tertiary is a giant with a mass of \hbmb M$_{\odot}$ and a radius of \hbrb R$_{\odot}$ in a \hbpab-day orbit with the binary. The system is again near-perfectly coplanar, with a mutual inclination of just \hbmutinc degrees.
In {\em TESS} Sector 33, the binary passes behind the giant tertiary, producing a distinctive ``head-and-shoulders'' eclipse that directly constrains the relative flux contributions and radii of the three stars. The tertiary dominates the system light in the {\em TESS} band at $\sim$95\%, with the primary and secondary of the eclipsing binary contributing $\sim$2.7\% and $\sim$2.3\% of the system light in the {\em TESS} band, respectively. Substantial radial velocity data from 48 TRES spectra, collected over more than four years, further constrained the system parameters by resolving the velocities of all three components.

We modeled the system using a comprehensive spectro-photodynamical model that simultaneously fits the {\em TESS} lightcurve, RVs, eclipse times, and the SED. The model incorporates MIST isochrone interpolation and SED modeling with an N-body dynamical simulation via {\sc REBOUND}.

The near-perfect coplanarity and compact configuration of TIC 295741342 are consistent with formation through disk fragmentation followed by inward orbital migration and gas dissipation. Evolutionary analysis using MIST tracks indicates that, for RGB-s, the tertiary will overflow its Roche lobe in \rgbrlof Myr at a radius of \rgbrlofrad R$_\odot$ as it ascends the RGB.  In HB-s, the the tertiary will ascend the AGB and overflow its Roche lobe in \hbrlof Myr at a radius of \hbrlofrad R$_{\odot}$.  In both cases, RLOF will be followed by either stable mass transfer to the binary or unstable common envelope evolution.

TIC 295741342 is one of only a handful of known triply-eclipsing triple star systems with a giant tertiary, and it has by far the lowest mutual inclination among them. The precise and recurring nature of the outer eclipse makes this system an excellent candidate for long-term monitoring and evolutionary studies. We predict that the midpoint of the next outer eclipse will occur on September 1, 2026 at 22:34:22 UTC, and we strongly encourage observations within a $\pm$3 day window around this date. Incorporating such an observation into future modeling would substantially reduce the uncertainty in the system parameters and enable accurate predictions of outer eclipses for years to come.


This paper includes data collected by the {\em TESS} mission, which are publicly available from the Mikulski Archive for Space Telescopes (MAST). The specific observations analyzed can be accessed via \dataset[doi:10.17909/nkwe-t219]{https://doi.org/10.17909/nkwe-t219}. Funding for the {\em TESS} mission is provided by NASA's Science Mission directorate.

The spectroscopic observations for this work were obtained with the assistance
of P.\ Berlind, M.\ Calkins, and G.\ Esquerdo. We thank them for their help.

Resources supporting this work were provided by the NASA High-End Computing (HEC) Program through the NASA Center for Climate Simulation (NCCS) at Goddard Space Flight Center.
We also made use of the Smithsonian High Performance Cluster (SI/HPC),
Smithsonian Institution (\url{https://doi.org/10.25572/SIHPC}).

V. B. K. acknowledges support from NASA grants 80NSSC22K0747 and 80NSSC23K0270

This work was partly supported by the HUN-REN Hungarian Research Network.
T.B. acknowledges the financial support of the Hungarian National Research,
Development and Innovation Office - NKFIH/OTKA Grant K-147131.

\facilities{
\emph{Gaia},
MAST,
TESS,
NCCS,
TRES,
}

\software{
{\tt Astropy} \citep{astropy2013,astropy2018}, 
{\tt IPython} \citep{ipython},
{\tt Keras} \citep{keras},
{\tt LcTools} \citep{2019arXiv191008034S,2021arXiv210310285S},
{\tt Lightkurve} \citep{lightkurve},
{\tt Matplotlib} \citep{matplotlib},
{\tt Mpi4py} \citep{mpi4py2008},
{\tt NumPy} \citep{numpy}, 
{\tt Pandas} \citep{pandas},
{\tt SciPy} \citep{scipy},
{\tt Tensorflow} \citep{tensorflow},
{\tt Tess-point} \citep{tess-point}
}

\bibliography{refs}{}
\bibliographystyle{aasjournalv7}

\end{document}